\documentclass[epj]{svjour}
\usepackage{graphics}
\usepackage{color}
\begin{document}
\title{Alpha-like correlations in $^{20}$Ne, comparison of quartetting wave function and THSR approaches}
\author{G.~R\"{o}pke\inst{1} \and C.~Xu\inst{2} \and B.~Zhou\inst{3,4} \and Z.~Z.~Ren\inst{5}  \and Y.~Funaki\inst{6}  \and H.~Horiuchi\inst{7}  \and M.~Lyu\inst{8}  \and A.~Tohsaki\inst{7}  \and T.~Yamada\inst{6} 
}                     
%
%
\institute{Institut f\"{u}r Physik, Universit\"{a}t Rostock, D-18051 Rostock, Germany 
\and School of Physics, Nanjing University, Nanjing 210093, China 
\and Key Laboratory of Nuclear Physics and Ion-Beam Application (MoE), Institute of Modern Physics, Fudan University, 200433, Shanghai, China
\and Shanghai Research Center for Theoretical Nuclear Physics, NSFC and Fudan University, 200438, Shanghai, China
\and School of Physics Science and Engineering, Tongji University, Shanghai 200092, China 
\and Laboratory of Physics, Kanto Gakuin University, Yokohama 236-8501, Japan 
\and Research Center for Nuclear Physics (RCNP), Osaka University, Osaka  567-0047, Japan 
\and College of Science, Nanjing University of Aeronautics and Astronautics, Nanjing, China 
}

\date{Received: date / Revised version: date}
%
\abstract{
$^{20}$Ne can be considered as a double-magic  $^{16}$O core nucleus surrounded by  four nucleons, the constituents of  an $\alpha$-like quartet.
Similar to other nuclei ($^{212}$Po, $^{104}$Ti, etc.) with a quartet on top of a double-magic core nucleus, significant $\alpha$-like correlations are expected.
Correlations in the ground state of $^{20}$Ne are investigated using different approaches. The quartetting wave function approach (QWFA) predicts a large $\alpha$-like cluster contribution 
near the surface of the nuclei. The Tohsaki-Horiuchi-Schuck-R\"opke (THSR) approach describes $\alpha$-like clustering in nuclear systems. The results of the QWFA in the Thomas-Fermi and shell-model approximation are compared with THSR calculations for the container model. Results for the $\alpha$ formation probability and the rms radii are shown.
\PACS{
      {PACS-key}{discribing text of that key}   \and
      {PACS-key}{discribing text of that key}
     } 
} 
\maketitle
\section{Introduction}

The liquid-drop model of nuclei, which can be considered as a simple version of a local density approach, reflects important properties of nuclear structure, for example the famous Bethe-Weizs{\"a}cker mass formula. Other properties such as the occurrence of magic numbers are explained by the shell model, which considers nucleonic quasiparticle states, and many properties of nuclei, including pairing, are studied in this approach, see e.g. the fundamental book by Ring and Schuck \cite{RingSchuck}.
However, the description of correlations, in particular of $\alpha$-like clusters in nuclei, requires going beyond the quasiparticle approach.

The nucleus $^{212}$Po shows a significant $\alpha$-like correlation in the skin region \cite{Po14,Xu16,Xu17}. 
 It can be assumed that it consists of a double-magic, relatively stable $^{208}$Pb core nucleus surrounded by an $\alpha$-like cluster.
 This $\alpha$-like quartet experiences a potential pocket for the center-of-mass (c.m.) motion outside a critical radius $r_{\rm cr}$ where it can exist as a quasi-bound state. 
 Its intrinsic structure is dissolved at smaller distances when the nucleon density of the core nucleus exceeds a critical value $n_{\rm cr}=0.03$ fm$^{-3}$. 
 The reason for this is the Pauli principle, which applies to the nucleons that form the $\alpha$ particle. Their mutual interaction is blocked in the dense medium of the nucleons of the core nucleus that occupy the Fermi sphere in momentum space. This is a consequence of the antisymmetrization of the full many-fermion wave function of the entire nucleonic system.
 The $\alpha$ particle, which is preformed in a near-surface pocket, can escape from the $^{212}$Po nucleus by tunneling. 
 The calculations were performed using the quartet wave function approach (QWFA).  
 It was found that the calculated $\alpha$ decay half-life agrees well with the observed data \cite{Xu16,Xu17}.

A similar behavior is expected for other nuclei consisting of a double magic core nucleus and an additional $\alpha$ cluster. 
 Calculations were performed for $^{104}$Te \cite{Yang20}.
 The observed half-life of $\alpha$ decay was successfully reproduced in QWFA for $^{104}$Te as well as for additional $\alpha$-decaying nuclei \cite{Yang21}. 
Improvements of the quartet model have been made in Refs. \cite{Bai19,Wang22}, see also \cite{Jin23,Li23}. 
Using QWFA, the influence of $\alpha$-like clustering in nuclei on the nuclear symmetry energy was analyzed in Ref. \cite{Yang23}.

Another nucleus, consisting of a double-magic core nucleus surrounded by an $\alpha$-like cluster, is $^{20}$Ne. In this work, we present calculations within QWFA and compare them with other approaches. 
A main result is the preformation fraction of $\alpha$-like clusters and the point rms radius which are determined by the wave function including correlations. We compare the Thomas-Fermi approximation with shell model calculations.  
 
A consistent description of quartetting ($\alpha$-like correlations) has recently been developed 
in the framework of the Tohsaki-Horiuchi-Schuck-R\"opke (THSR) approach \cite{THSR}.  This approach provides an excellent 
description of low-density 4$n$ nuclei such as $^8$Be, the Hoyle state of $^{12}$C and excited states 
of $^{16}$O, but has also been applied to more complex nuclei such as $^{20}$Ne \cite{Bo12,Bo13,Bo14} as well as
4$n$ nuclei with additional nucleons \cite{Lyu,Lyu1}. Recently, calculations for $^{20}$Ne were performed  by Bo {\it et al.} \cite{Bo23} using  the two-parameter container model. 
A review on microscopic clustering in light nuclei was presented in Ref. \cite{Freer18}. 

In this work, we also compare the two approaches. 
Heavy nuclei with a large number of nucleons like $^{212}$Po are not yet computable with the THSR approach. 
The QWFA provides better results for heavier nuclei where a mean-field approach for the core nucleus is more justified.
For
$^{20}$Ne, both approaches are feasible. The comparison of the results for the quartetting wave function approach
and THSR calculations allows a better understanding of the description of correlations in nuclear systems.

We study the c.m. motion of a quartet $\{ n_\uparrow, n_\downarrow, p_\uparrow, p_\downarrow \}$ 
as a new collective degree of freedom and compare the wave functions for both approaches, the QWFA and the THSR approach.
Instead of an uncorrelated Fermi gas model for the cluster environment, an improvement of the quartet wave function approach is investigated to achieve a consistent description of cluster formation in a clustered medium. 

After a brief explanation of the QWFA in Sec. \ref{sec:2}, we  carry out calculations using the Thomas-Fermi approach in Sec. \ref{sec:3}. Calculations with the shell model are shown in Sec. \ref{sec:shell}. 
 Comparisons with the THSR approach are discussed in Sec. \ref{sec:THSR}, and concluding remarks are made in Sec. \ref{sec:Con}.

\section{The quartet wave equation}
\label{sec:2}

The treatment of the interacting many-nucleon system requires some approximations 
which can be obtained in a consistent way from a Green's function approach. 
The quartetting wave function approach \cite{Po14,wir} considers the wave function 
$\Psi({\bf r}_1 {\bf r}_2 {\bf r}_3{\bf r}_4)$ of the quartet 
(spin and isospin variables are dropped) which obeys the in-medium wave equation
\begin{eqnarray}
&&[E_4\!-\!\hat h_1\! -\!\hat h_2\!-\! \hat h_3\! - \!\hat h_4]\Psi({\bf r}_1,{\bf r}_2,{\bf r}_3,{\bf r}_4)
\nonumber \\ &&
=\int \!\! d^3 {\bf r}_1'\,d^3 {\bf r}_2' \langle {\bf r}_1{\bf r}_2|\hat B(1,2) \,\,\hat V_{N-N}| {\bf r}_1'{\bf r}_2'\rangle
\Psi({\bf r}_1',{\bf r}_2',{\bf r}_3,{\bf r}_4)\nonumber \\ && +
\int d^3 {\bf r}_1'\,\,d^3{\bf r}_3'  \langle {\bf r}_1{\bf r}_3|\hat B (1,3) \,\,\hat V_{N-N}|
{\bf r}_1'{\bf r}_3'\rangle \Psi({\bf r}_1',{\bf r}_2,{\bf r}_3',{\bf r}_4)
\nonumber \\ &&
+ {\rm four\,\, further \,\,permutations,}
\label{15}
\end{eqnarray}
with the single-quasiparticle Hamiltonian (single-nucleon shell states $|n,\nu \rangle$)
\begin{equation}
\label{spPauli}
 \hat h_i=\frac{\hbar^2 \hat p_i^2}{2m}+ [1 - \hat f_{\nu_i}]\, V_{\nu_i}^{\rm mf}(\hat r),\qquad
\hat f_\nu =\sum_n^{{\rm occ.}}| n,\nu \rangle \langle n,\nu |
\end{equation}
denotes the phase space which, according to the Pauli principle, cannot be used for an interaction process of a nucleon
with an intrinsic quantum state $\nu=\sigma,\,\tau$.
In addition to the nucleon-nucleon potential $\hat V_{N-N}$, the nucleon-nucleon interaction terms also contain the 
 blocking operator $ \hat B(1,2)=[1-\hat f_1-\hat f_2]$ for the first term on the r.h.s. of Eq. (\ref{15}), and corresponding expressions for the other 5 terms.
The mean-field potential $V_{\nu_i}^{\rm mf}(\hat r)$ contains  the strong core-nucleon interaction 
$V^{\rm ext}( r)$ as well as  the Coulomb potential of the core nucleus. It is considered as an external potential. 
The Pauli blocking terms, which are given by the occupation numbers $\hat f_\nu $, are not easy to treat 
as will be explained below. The mean-field approach treats the motion within the cluster independent of the motion in the surrounding medium,
and neglects any correlations between the two.  If such further correlations exist, clusters with a larger number of nucleons are formed.
This concept is known from the shell model at the one-particle level, for pairing at the two-particle level. 
We first discuss here the motion of four nucleons under the influence of an external potential.

A main aspect of the cluster approach is the introduction of the center-of-mass (c.m.) motion $\bf R$ 
as new collective degree of freedom, and ${\bf s}_j=\{\bf S,s,s'\}$ for the intrinsic motion
(Jacobi-Moshinsky coordinates).
As shown in \cite{Po14}, the normalized quartet wave function $\Phi({\bf R},{\bf s}_j) $,
\begin{equation}
\int d^3R\,\int d^9s_j\,|\Phi({\bf R},{\bf s}_j)|^2 =1,
\end{equation}
can be decomposed in a unique way (up to a phase factor),
\begin{equation}
\label{4}
\Phi({\bf R},{\bf s}_j)=\varphi^{{\rm intr}}({\bf s}_j,{\bf R})\,\Psi^{\rm c.m.}({\bf R})
\end{equation}
with the individual normalizations 
 \begin{equation}
 \label{normS}
\int d^3R\,|\Psi^{\rm c.m.}({\bf R})|^2=1
\,, {\rm and} 
\int d^9s_j |\varphi^{{\rm intr}}({\bf s}_j,{\bf R})|^2=1
\end{equation}
for arbitrary ${\bf R}$.

The Hamiltonian of a four-nucleon cluster 
can be written as 
\begin{eqnarray}
H&=&\left(-\frac{\hbar^2}{8m} \nabla_R^2+T[\nabla_{s_j}]\right)\delta^3({\bf R}-{\bf R}')\delta^3({\bf s}_j-{\bf s}'_j)
\nonumber \\&&
+V({\bf R},{\bf s}_j;{\bf R}',{\bf s}'_j)
\end{eqnarray}
with the kinetic energy of the c.m. motion of the cluster, and the kinetic energy  $T[\nabla_{s_j}]$ of the internal motion. 
The interaction $V({\bf R},{\bf s}_j;{\bf R}',{\bf s}'_j)$  contains the mutual interaction $V_{ij}({\bf r}_i,{\bf r}_j,{\bf r}'_i,{\bf r}'_j)$ 
between the quartet particles 
as well as the 
interaction with an external potential (for instance, the mean-field potential of the core nucleus), with strict fulfillment of the Pauli principle. 

For the c.m. motion we have the wave equation
\begin{eqnarray}
\label{9}
&&-\frac{\hbar^2}{8m} \nabla_R^2\Psi^{\rm c.m.}({\bf R})-\frac{\hbar^2}{4m}\int d^9s_j 
\nonumber \\&& \times
\varphi^{{\rm intr},*}({\bf s}_j,{\bf R}) 
[\nabla_R \varphi^{{\rm intr}}({\bf s}_j,{\bf R})][\nabla_R\Psi^{\rm c.m.}({\bf R})]-
\nonumber \\&&
-\frac{\hbar^2}{8m}\int\!\! d^9s_j \varphi^{{\rm intr},*}({\bf s}_j,{\bf R}) 
[ \nabla_R^2 \varphi^{{\rm intr}}({\bf s}_j,{\bf R})] \Psi^{\rm c.m.}({\bf R})
\nonumber \\ &&
+\!\!\int \!\! d^3R'\,W({\bf R},{\bf R}')  ,\Psi^{\rm c.m.}({\bf R}')\!=\!E\,\Psi^{\rm c.m.}({\bf R}),\nonumber 
\end{eqnarray}
with the c.m. potential
\begin{eqnarray}
\label{9c}
&&W({\bf R},{\bf R}')
=\int d^9s_j\,d^9s'_j\,\varphi^{{\rm intr},*}({\bf s}_j,{\bf R}) \left[T[\nabla_{s_j}]
\right. \\&&\left.
\!\!\!\!\!\!\!\!\!\!\times
\delta^3({\bf R}-{\bf R}')\delta^9({\bf s}_j-{\bf s}'_j)
+V({\bf R},{\bf s}_j;{\bf R}',{\bf s}'_j)\right]
\varphi^{{\rm intr}}({\bf s}'_j,{\bf R}').\nonumber
\end{eqnarray}
For the intrinsic motion we find the wave equation
\begin{eqnarray}
\label{10}
&&-\frac{\hbar^2}{4m}  \Psi^{\rm c.m.*}({\bf R}) [\nabla_R\Psi^{\rm c.m.}({\bf R})]
[\nabla_R \varphi^{{\rm intr}}({\bf s}_j,{\bf R})]
\nonumber \\&&
-\frac{\hbar^2}{8m}  |\Psi^{\rm c.m.}({\bf R})|^2
\nabla_R^2 \varphi^{{\rm intr}}({\bf s}_j,{\bf R})
\nonumber \\ &&
+\int d^3R'\,d^9s'_j\, \Psi^{\rm c.m.*}({\bf R}) \left[T[\nabla_{s_j}]
\delta^3({\bf R}-{\bf R}')\delta^9({\bf s}_j-{\bf s}'_j)\right.\nonumber \\&& \left.
+V({\bf R},{\bf s}_j;{\bf R}',{\bf s}'_j)\right]
\Psi^{\rm c.m.}({\bf R}')\varphi^{{\rm intr}}({\bf s}'_j,{\bf R}')
\nonumber \\&&
=F({\bf R}) \varphi^{{\rm intr}}({\bf s}_j,{\bf R})\,.
\end{eqnarray}

Both the c.m. and intrinsic Schr\"odinger
equations, Eqs. (\ref{9}) and (\ref{10}), respectively,  are coupled by contributions containing the expression
$\nabla_R \varphi^{{\rm intr}}({\bf s}_j,{\bf R})$. This expression vanishes in homogeneous matter, 
and we recover the in-medium Schr\"odinger equation for $\alpha$ clusters in matter without external potential. 
Then, the eigenvalue $F({\bf R})$ of Eq. (\ref{10}) represents the bound state energy of the $\alpha$ particle 
which is shifted in dense matter because of Pauli blocking.

The contribution of the gradient terms was recently investigated by Yang et al. \cite{Yang23}.
It can be shown that the second term of Eq. (\ref{9}) vanishes. 
In the present work, we neglect the contributions of the gradient terms.
This corresponds to a local density approximation, as is often used in many-body theories.

\section{Quartets in nuclei in Thomas-Fermi approximation}
\label{sec:3}

\subsection{Mean field for the c.m. motion}

We would like to emphasize that in general non-local interactions are possible. In particular, the Pauli blocking considered in the following is non-local. 
To simplify the calculations, local approximations are often used,
\begin{eqnarray}
\label{eq:10}
&&W({\bf R},{\bf R}')\approx W({\bf R})\delta^3({\bf R}-{\bf R}'),
\nonumber \\&&
 W({\bf R})=W^{\rm ext}({\bf R})+W^{\rm intr}({\bf R}).
\end{eqnarray}
$W^{\rm ext}({\bf R})=W^{\rm mf}({\bf R})$ is the contribution of external 
potentials, here the mean field of the core nucleons.
The interaction within the cluster according  Eq. (\ref{10}) gives the contribution $W^{\rm intr}({\bf R})$.
We give a short description, for details see Refs. \cite{wir,R17,Ro18}.

If we know the nucleon densities of the core nucleus, the mean fields can be easily calculated.
The mean-field contribution $W^{\rm mf}({\bf R})$ is obtained by double folding the density distribution of the core 
nucleus and the intrinsic density distribution of the quartet at c.m. position $\bf R$ with the interaction potential of the nucleons. 
An $\alpha$-like Gaussian density distribution was assumed for the bound quartet. 

For the Coulomb interaction we calculate the double-folding potential
\begin{equation}
\label{VCoul}
V^{\rm Coul}_{\alpha - {\rm O}}( R) = \int d^3 r_1 \int d^3 r_2 \rho_{{\rm O}} ({\bf r}_1) \rho_\alpha ({\bf r}_2) 
\frac{e^2}{|{\bf R}-{\bf r}_1+{\bf r}_2|}\,.
\end{equation}
The charge density of the $\alpha$ nucleus according to 
\begin{equation}
\label{nqalpha}
 \rho_\alpha( r)=0.2114\,\,{\rm fm}^{-3} \,e^{-0.7024\,\, r^2/{\rm fm}^2} 
\end{equation}
reproduces the measured rms point radius 1.45 fm. For the density distribution
of $^{16}$O, the expression \cite{Qu2011}
\begin{equation}
\label{Qu}
n^{\rm WS}_{B,{\rm O}}( r) = \frac{0.168 \,{\rm fm}^{-3}}{1+e^{(r/{\rm fm}-2.6)/0.45}}
\end{equation}
was given which reproduces the rms point radius 2.6201 fm, or Gaussians \cite{wir}.
The convolution integral (\ref{VCoul}) is easily evaluated in Fourier representation and gives 
for the parameter values considered here \cite{wir}
\begin{eqnarray}
\label{coulao}
&& V^{\rm Coul}_{\alpha - {\rm O}}( R)=\frac{16 \times 1.44}{R} \,{\rm MeV\,\, fm}  \\&& \nonumber \times\left[{\rm Erf}(0.7683\,\, R/{\rm fm})-0.9097 \,\,(R/{\rm fm})\,\,e^{-0.2274\,\,R^2/{\rm fm}^2}\right]\,.
\end{eqnarray}

For the nucleon-nucleon contribution to the mean field, 
a parametrized effective nucleon interaction (distance $s$)
\begin{equation}
\label{VNN}
 V_{N-N}(s/{\rm fm})=c\, \exp(-4 s)/(4 s)-d\, \exp(-2.5 s)/(2.5 s)
\end{equation}
can be used which is motivated by the M3Y interaction \cite{M3YReview}, 
$s$ denotes the distance of nucleons. The parameters $c, d$ are adapted 
to reproduce known data, see \cite{Po14,Xu16,Xu17}
for the case of a lead core nucleus. For the oxygen core nucleus,
parameter values $c,d$ are given below in Eq. (\ref{cd}).
As also known from other mean-field approaches, we fit the mean field parameter to measured data.
The nucleonic contribution $V^{\rm N-N}_{\alpha - {\rm O}}( R)$ to the mean field is calculated in analogy to Eq. (\ref{VCoul}) replacing the Coulomb interaction by the nucleon interaction (\ref{VNN}).
With both contributions, the mean-field part of the cluster potential is 
\begin{equation}
\label{ext}
W^{\rm ext}({\bf R})=W^{\rm mf}({\bf R})=V^{\rm Coul}_{\alpha - {\rm O}}( R)+V^{\rm N-N}_{\alpha - {\rm O}}( R). 
\end{equation}

The local approximation $W^{\rm intr}({\bf R})$, Eq. (\ref{eq:10}), for the intrinsic contribution 
to the effective c.m. potential is more involved. 
It contains the binding energy of the cluster taking into account the Pauli blocking 
of the surrounding matter. The local density approximation neglects any gradient terms 
so that homogeneous-matter results can be used.

The intrinsic wave equation (\ref{10})  describes in the zero density limit
the formation of an $\alpha$ cluster with binding energy $B_\alpha= 28.3$
MeV. In homogeneous matter, the binding energy is reduced
due to Pauli blocking. The shift of the binding energy is determined by the baryon
density $n_B=n_n+n_p$ and the asymmetry $\delta= 2 n_p/n_B-1$. For the c.m. momentum ${\bf P}=0$, the Pauli
blocking term depends on the  baryon density $n_B$ \cite{Po14,wir} as
\begin{eqnarray}    
\label{WPauli}
 W^{\rm Pauli}(n_B,\delta)&\approx& 4515.9\, {\rm MeV\, fm}^3 n_B \nonumber \\ &&-100935\, {\rm MeV\, fm}^6 n_B^2 (1+\delta^2)\nonumber \\ &&+1202538\, {\rm MeV\, fm}^9 n_B^3(1+3 \delta^2)\,.
\end{eqnarray}
This approximation  formula applies to the density  
range  $n_B \le n_{\rm crit}= 0.02917$ fm$^{-3}$.
In particular, the bound state is dissolved and merges with the continuum
of the scattering states at the critical density $n_{\rm crit}$ 
(introduced as Mott density). 
A more detailed discussion of this ansatz for the Pauli blocking term will follow below, see section \ref{sec:shell}.
For the intrinsic wave function of the quartet, we can assume an $\alpha$-like Gaussian function
to describe the bound state. The width parameter of the free $\alpha$ particle is only 
weakly changed when it approaches  the critical density, see Ref. \cite{Po14}.

Below  the critical density, $n_B \le n_{\rm crit}$, the intrinsic potential
\begin{equation}
\label{WeffR}
W^{\rm intr}({\bf R})=-B_\alpha+ W^{\rm Pauli}[n_B({\bf R})], \qquad n_B \le n_{\rm crit}
\end{equation}
results in a local density approximation. 
The intrinsic energy of the quartet for densities above the critical density is a minimum if all four nucleons are at the Fermi energy  (ideal Fermi gas),
for symmetric matter and  $n_B \ge n_{\rm crit}$
\begin{equation}
W^{\rm intr}({\bf R})=4 E_F[n_B({\bf R})], 
\end{equation}
with
\begin{equation}
E_F(n_B)=(\hbar^2/2m) (3 \pi^2n_B/2)^{2/3}.
\end{equation}

\subsection{Thomas-Fermi rule and results for $^{20}$Ne in local density approximation}
\label{TFR}

\begin{figure}
\resizebox{0.5\textwidth}{!}{%
\includegraphics{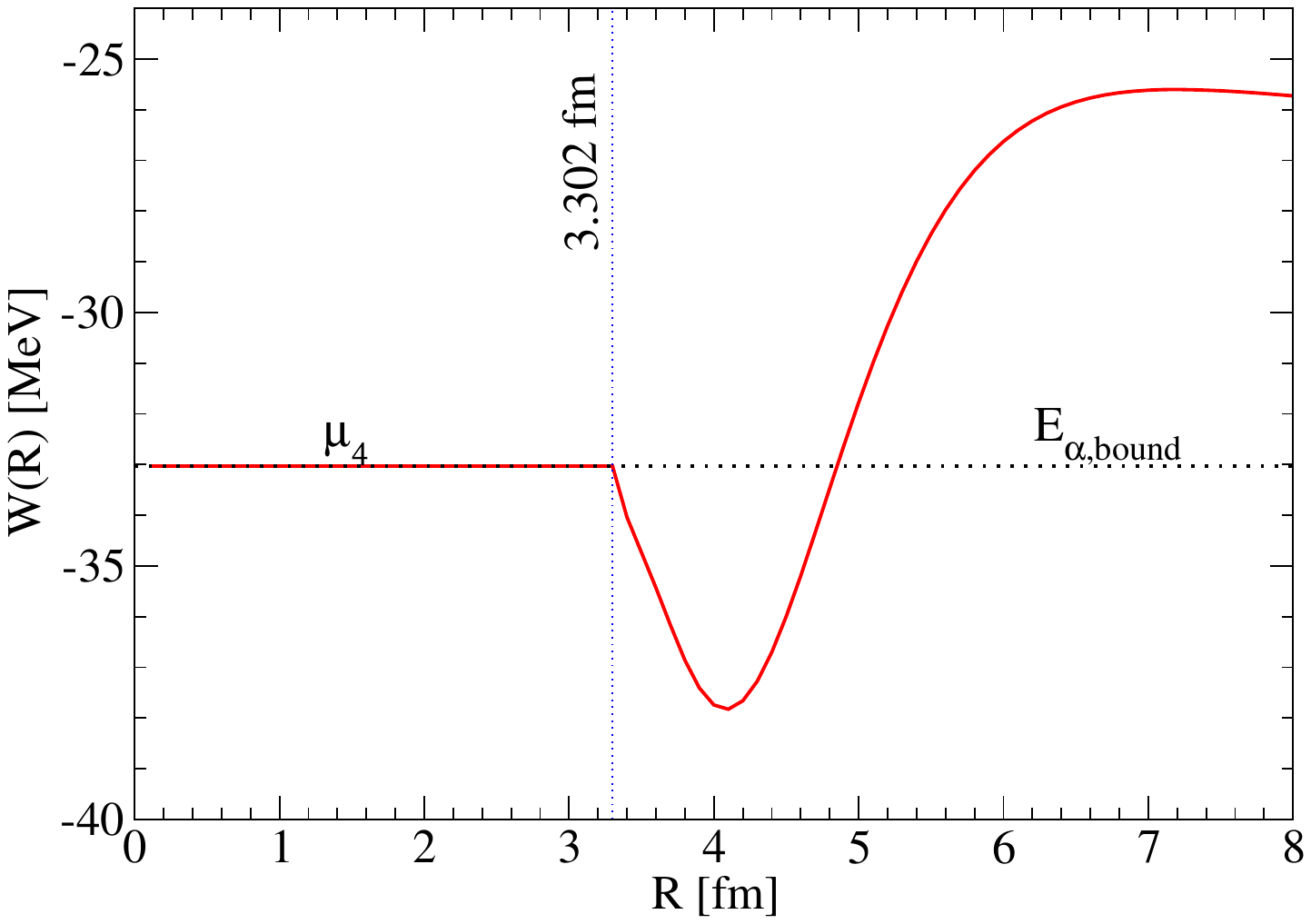}
}
  \caption{Effective potential $ W^{\rm TF}( R) $ for the center of mass motion of the quartet on top of $^{16}$O. 
The Thomas-Fermi model has been used. The formation of a pocket is shown.}
\label{fig:2}
\end{figure}

The quartetting wave function approach for  $^{20}$Ne in local density approximation 
has been considered in Ref.~\cite{wir}. We are presenting some results for the effective potential $W( {\bf R})$
and the wave function $\psi({\bf R})$, see Fig. \ref{fig:2}. To this purpose, we use empirical data from the nuclei involved.

The mean-field contribution $W^{\rm ext}( R)$ (\ref{ext})
is given by the double-folding Coulomb and $N-N$ potentials. 
Empirical values for the densities of the $\alpha$ particle (\ref{nqalpha}) 
and the $^{16}$O core nucleus (\ref{Qu})
are known from scattering experiments, such as the rms radii, so that the Coulomb interaction 
$V^{\rm Coul}_{\alpha - {\rm O}}( R)$ (\ref{coulao}) as well as the  nucleon-nucleon interaction 
$V^{\rm N-N}_{\alpha - {\rm O}}( R)$ can be calculated. 

With respect to $W^{\rm intr}({\bf R})$, the local density approximation is 
also very simple:
There are two regions separated by the critical radius $r_{\rm crit} =3.302$ fm in which the density 
of the $^{16}$O core nucleus (\ref{Qu}) has the critical value 
$n_B(r_{\rm crit})= n_{\rm crit}= 0.02917$ fm$^{-3}$. 
We obtain $-B_\alpha+W^{\rm Pauli}[n_B(r_{\rm crit})]=4 E_F[n_B(r_{\rm crit})]$, and the bound state merges 
with the continuum of scattering states. 

For $R > r_{\rm crit}$, the intrinsic part $W^{\rm intr}( R)$ 
contains the bound state energy -28.3 MeV of the free $\alpha$ particle, which is shifted due to Pauli blocking. 
At $r_{\rm crit}$ the bound state merges with the continuum, so that we have the condition (symmetric matter)
\begin{equation}
 W(r_{\rm crit})= W^{\rm ext}(r_{\rm crit})+4 E_F(n_{\rm crit}) =\mu_4,
\end{equation}
the intrinsic wave function changes from a bound state to four uncorrelated quasiparticles on top of the Fermi sphere (the states below the Fermi energy are already occupied).

For $R < r_{\rm crit}$, the Fermi energy $4 E_F[n( R)]$ appears in addition to the mean-field contribution $W^{\rm ext}( R)$. 
In the Thomas-Fermi model, for a given potential $W^{\rm ext}( R)$ 
the density is determined by the condition that $W^{\rm ext}(R )+4 E_F[n_B( R)]$ remains a constant, 
here $\mu_4$.
We find the effective potential $W^{\rm TF}( R)$, which is continuous but has a kink at $r_{\rm crit}$.
It is an advantage of the Thomas-Fermi model that the condition $W^{\rm TF}( R)=\mu_4=$ const applies 
to the entire range $R < r_{\rm crit}$, 
independently of the shape of the mean-field potential $W^{\rm ext}( R)$ and the corresponding density distribution. 
We analyze this property in the following section.

While the Coulomb part of the external potential as well as the intrinsic part of the effective potential 
$W^{\rm TF}( R)$ are fixed, the two parameters $c,d$ in Eq. (\ref{VNN}) for the $N-N$ part of the external potential 
can be adjusted such 
that measured data are reproduced. In the case of heavy nuclei that are $\alpha$ emitters, such as $^{212}$Po \cite{Po14}, two conditions can be formulated:\\ 
i) For $\alpha$ emitters, the normalized solution of the c.m. wave equation (neglecting the decay) gives the energy eigenvalue $E_\alpha=E_{\rm tunnel}$. 
This eigenvalue should correspond to the measured energy after decay, which is given by the $Q$ value.\\
ii) This value $E_{\rm tunnel}$ should coincide with the value $\mu_4$. 
In the context of the  local density approach, this is the value that the four nucleons must have in order to implement them into the core nucleus. We denote this condition 
\begin{equation}
\label{eq:TFR}
E_\alpha=\mu_4
\end{equation}
 as the Thomas-Fermi rule \cite{wir}.
With both conditions, the parameter $c,d$ for the double folding $N - N$ interaction potential are found, 
and values for the preformation factor and the half-life of the $\alpha$ decay were determined for heavy nuclei, see Ref. \cite{Po14,Xu16,Xu17}, where further discussions were made.

In contrast to the $\alpha$ decay of $^{212}$Po where the $Q$ value can be used to estimate the chemical potential $\mu_4$ \cite{Po14}, the  nucleus $^{20}$Ne is stable. However, 
we can use the additional bonding in the transition from  $^{16}$O ($B_{^{16}{\rm O}}=127.66$ MeV) to  $^{20}$Ne ($B_{^{20}{\rm Ne}}=160.645$ MeV) 
by adding the four nucleons. The difference fixes  
the position of the in-core effective potential $\mu_4 =B_{^{16}{\rm O}}-B_{^{20}{\rm Ne}}=-33.0$ MeV. 

Another condition is that  the solution of the Schr{\"o}dinger equation for the four-nucleon c.m. motion in the effective potential $W( R)$ gives the energy eigenvalue 
$E_{\alpha, {\rm bound}}$ at this value -33 MeV, so that the energy eigenvalue of the $\alpha$-like cluster coincides with the  
Fermi energy $\mu_4 $ (the Thomas-Fermi rule, see also the discussion in Ref. \cite{Xu16}). 
Both conditions are used  to fix the parameters $c, d$.  The values 
\begin{equation}
\label{cd}
 c=4650\,\,{\rm  MeV \,\,\,\, and}\,\,\, d=1900\,\,{\rm MeV} 
\end{equation}
have been found \cite{wir}.

The resulting effective potential $ W^{\rm TF}( R) $ (\ref{WeffR}) 
for the center of mass motion of the quartet is shown in Fig. \ref{fig:2}. 
One can see the formation of a pocket near the surface caused by the formation of an $\alpha$-like cluster. The sharp kink at the critical radius 
$r_{\rm crit}=3.302$ fm is a consequence of the local approximation for the Pauli blocking term. A smooth behavior is expected if the finite extension of the 
$\alpha$-like cluster is taken into account so that the kink generated by the local density approximation is smeared out.

\begin{figure}
\resizebox{0.5\textwidth}{!}{%
\includegraphics{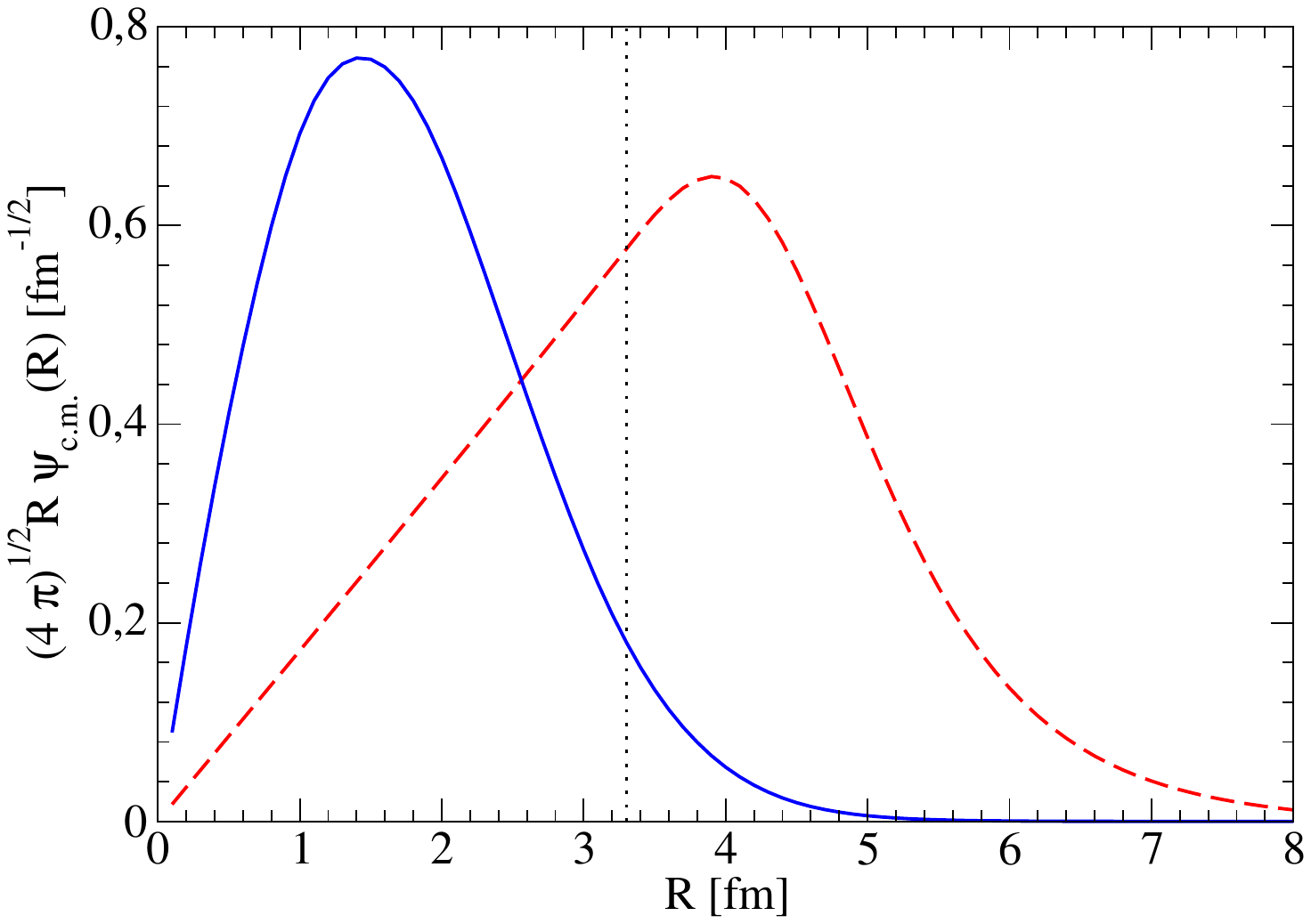}
}
  \caption{Wave function for the c.m. motion of the quartet. A prefactor $(4 \pi)^{1/2} R$ is introduced so that the integral over $R$ of the squared quantity is normalized to 1. The solution for the Thomas-Fermi model $\psi^{\rm TF}_{\rm c.m.}( R)$ (red, dashed) is compared with the non-interacting shell-model calculation $\psi_{2s^4}( R)$ (blue). The shift of the maximum is caused by the formation of a pocket, see Fig.~\ref{fig:2}. Dotted line: $r_{\rm crit}=3.302$ fm.}
\label{fig:3}
\end{figure}

The wave function for the quartet center-of-mass motion $\psi^{\rm TF}_{\rm c.m.}( R)$ is found as a solution of the Schr\"odinger equation, mass $4 m$, 
with the potential $ W^{\rm TF}( R) $. The energy eigenvalue is -33.0 MeV. A graph of $(4 \pi)^{1/2} R\, \psi^{\rm TF}_{\rm c.m.}( R)$ is shown in Fig. \ref{fig:3}. As a result, in Ref. \cite{wir} the rms point radius 2.864 fm for  $^{20}$Ne
was calculated
which is in good agreement with the experimental rms point radius of 2.87 fm. 
The normalization is $4 \pi \int_0^\infty r^2 \psi^2_{\rm c.m.}(r) dr =1$. Integrating from 0 to $r_{\rm crit}=3.302$ fm, the part of the quartet 
where the internal structure is the product of free states, comes out at 0.3612. The remaining part where the internal structure is given by an $\alpha$-like
bound state is 0.6388.

A further discussion of optical model description and double-folding potential is given in App. \ref{app:1}.
Note that the standard approaches of optical model potentials have a diverging repulsive potential below $r_{\rm crit}$.

\subsection{Discussion of the Thomas-Fermi rule $E_\alpha = \mu_4$}

The condition  $E_\alpha = \mu_4$ (\ref{eq:TFR}) is a consequence of the Thomas-Fermi model, which applies to infinite matter: an additional nucleon with given spin and isospin can be introduced at the corresponding chemical potential $\mu_{\sigma, \tau}$. At zero temperature, this coincides with the corresponding Fermi energy (plus the potential energy). For finite system such as nuclei, 
the energy levels of the single-nucleon states are discrete.  If  we add a nucleon to the core nucleus  in which all the single-nucleon states below a certain energy are occupied, the next free single-nucleon state  that  is free has a distance to the chemical potential. This means, under these considerations, the quartet cannot be introduced at $\mu_4$ 
but at a higher value $E_\alpha > \mu_4$ which is now a new parameter. This aspect has been worked out already  in \cite{Xu17}. 
We do the same here for $^{20}$Ne.

We compare our calculations with values for $^{212}$Po. 
The $\alpha$ decay energy $Q_\alpha$ was introduced as the difference between the binding energy of the mother nucleus ($^{212}$Po)
 and the binding energies of the daughter nuclei ($^{208}$Pb and $\alpha$). Similarly, we have -4.73 MeV, so that the energy eigenvalue 
 of the Schr\"odingier equation comes out as $E^0_\alpha-Q_\alpha=-28.3-4.73$ MeV=-33.03 MeV.
 As a second condition, we used the results for $^{212}$Po. If $d=3415.56$ remains the same, the given energy eigenvalue is 
 of the Schr\"odingier equation is reproduced with $c=10623$. This results in the value $\mu_4=-32.388$ MeV and $P_\alpha=0.72$
 follow. If we take $c=11032$ from Po, we get $d=3513.46$ as well as $\mu_4=-32.12$ MeV and $P_\alpha=0.74$.
  
 We reproduce a large preformation factor $P_\alpha$ in both cases. In contrast to the Thomas-Fermi model, 
 the condition  $E_\alpha = \mu_4$ is not valid. The value of $\mu_4$ is not below $E_\alpha$ as expected from the shell model 
 consideration, but $E_\alpha < \mu_4$. This means that it is energetically more favorable for the nucleus to form correlated quartets
 instead  of remaining in uncorrelated single-nucleon (shell model) states. This will be seen from the THSR calculations, in which
 the core nucleus $^{16}$O  also shows $\alpha$-like correlations.

\section{Shell model calculations}
\label{sec:shell}

\subsection{Comparison with the harmonic oscillator model}
\label{sec:HO}

The local density approximation (Thomas-Fermi model) is not able to describe the 
nuclear structure of the core nucleus. In particular, the Thomas-Fermi rule must be replaced 
by a more microscopic approach, see \cite{Po14,Xu16,Xu17}. However, the behavior of the effective c.m. potential 
which remains  almost constant within the core nucleus, is also interesting in the case that shell model states are used.
 A first attempt was made in Ref. \cite{wir} 
with harmonic oscillator states. 
The results of the simple Thomas-Fermi model, 
in particular the approximate constancy of the c.m. quartetting potential in the core nucleus and the Thomas-Fermi rule, can be verified. 
However, the harmonic oscillator potential is not realistic 
for nuclei, especially near the surface of the core nucleus where $\alpha$-like quartets are formed.
We present here calculations with more realistic potentials (units MeV, fm), see also \cite{Mirea}. The intrinsic nucleon-nucleon interaction 
$W^{\rm intr}( R)$, which is suppressed due to Pauli blocking, is not considered in this section \ref{sec:HO}.

\begin{figure}
\resizebox{0.5\textwidth}{!}{%
  \includegraphics{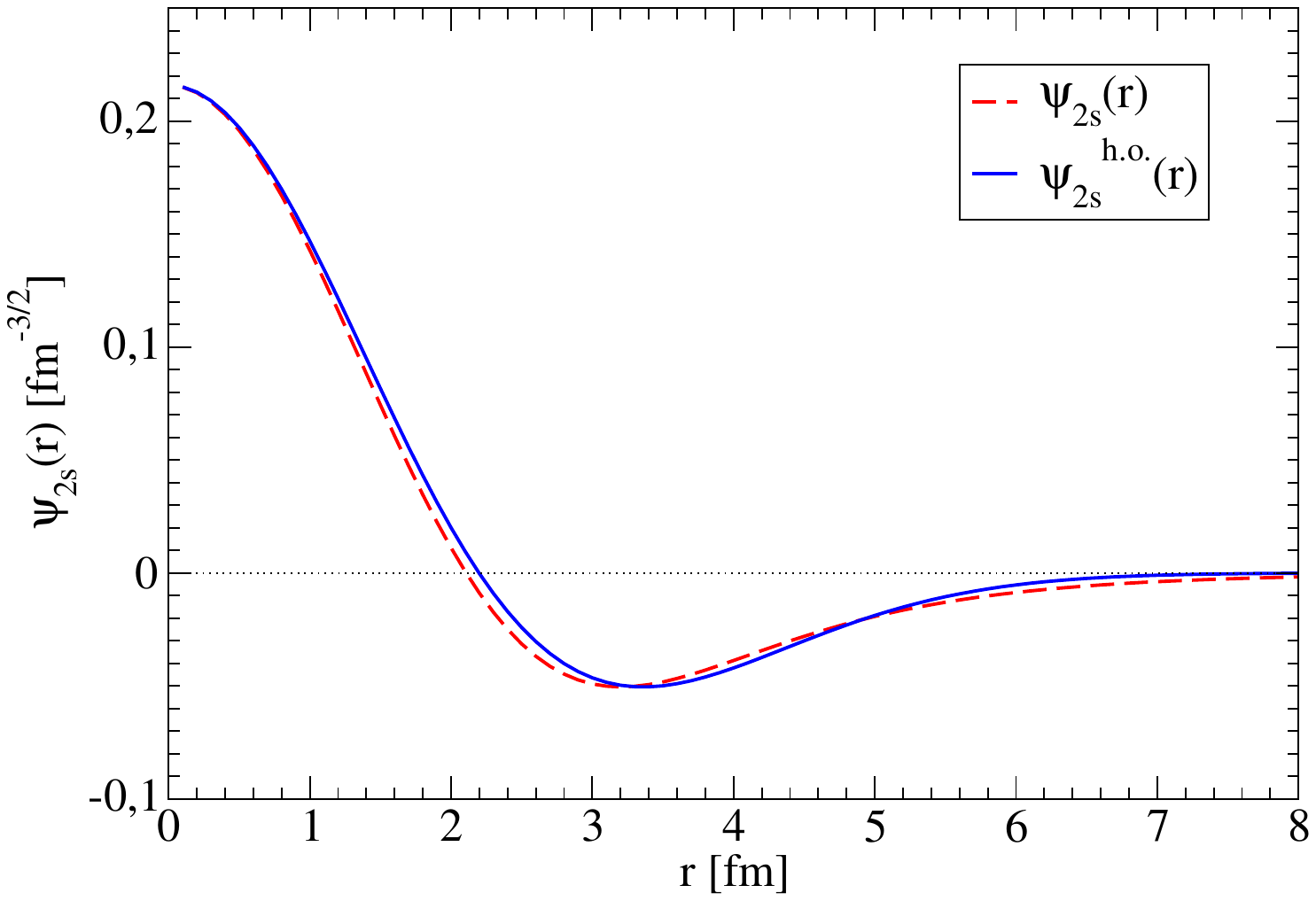}\\
\includegraphics{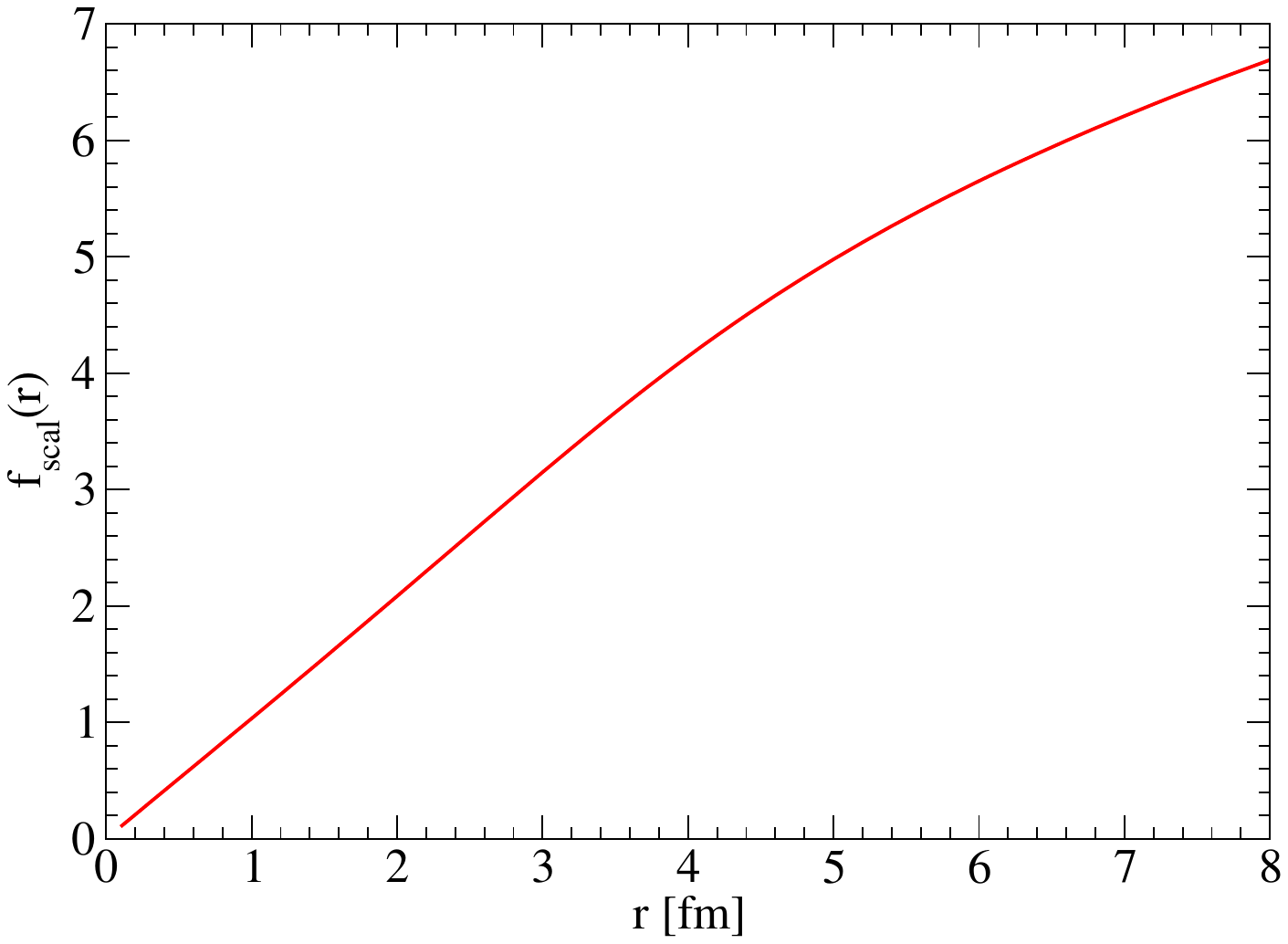}
}
 \caption{Normalized wave function $\psi_{2s}(r )$ for the Woods-Saxon potential (\ref{WS2s}). For comparison, the harmonic oscillator wave function $\psi_{2s}^{\rm HO}(r)$ is also given, where the potential parameter $a^{\rm HO}$ is chosen so that 
$\psi_{2s}(0)$ coincides. The scaling function $f_{\rm scal}(r )$ give full coincidence of both wave functions.}
\label{Fig:psi}
\end{figure}

A more systematic way to find a suitable simple basis of single-particle states
is to use the Woods-Saxon potential \cite{WS} for $Z=N$, see Ref. \cite{Ro18},
\begin{equation}
\label{WS2s}
 V_{\rm WS}(r)=\frac{V_0 (1+3\kappa/A)}{1+\exp[(r-R_0 A^{1/3})/a]}
\end{equation}
with $V_0=-52.06$ MeV, $\kappa=0.639$, $R_0=1.26$ fm, $a=0.662$ fm. 
The normalized solution $\psi_{2s}(r )$ for the 2$s$ state is shown in Fig. \ref{Fig:psi}, eigenvalue $E_{2s}=-9.162$ MeV. For comparison,
the harmonic oscillator wave function
\begin{equation}
 \psi_{2s}^{\rm HO}(r )=-\left(\frac{a^{\rm HO}}{\pi}\right)^{3/4} e^{-a^{\rm HO} r^2/2}\left(a^{\rm HO} r^2-\frac{3}{2}\right) \left(\frac{2}{3}\right)^{1/2},
\end{equation}
is also shown, where the parameter $a^{\rm HO}=0.31047$ fm is chosen so that 
the values coincide at $r=0$. A scaling of the $r$-axis is considered to make both coincide, 
$\psi_{2s}^{\rm HO}(r' )=\psi_{2s}(r )/(1+0.0024719\, r)$. (The amplitude correction is necessary to reproduce the correct value of the minimum).
This defines the relationship  $r'=f_{\rm scal}(r )$ shown in Fig. \ref{Fig:psi}.

Neglecting any intrinsic interaction, the  2$s$ wave functions can be used to construct the quartet  wave function
\begin{equation}
 \Phi_{2s^4}({\bf R, S, s, s}')=\psi_{2s}({\bf r}_{n,\uparrow})\,\psi_{2s}({\bf r}_{n,\downarrow})\,
\psi_{2s}({\bf r}_{p,\uparrow})\,\psi_{2s}({\bf r}_{p,\downarrow}).
\end{equation}
The wave function for the c.o.m. motion follows as (Jacobi-Moshinsky coordinates ${\bf R, S, s, s}'$ \cite{wir})
\begin{equation}
\label{Phi1s}
\psi_{2s^4}({\bf R})=\left[\int d^3Sd^3sd^3s'|\Phi_{2s^4}({\bf R, S, s, s}')|^2\right]^{1/2}\,.
\end{equation}

The evaluation of the 9-fold integral in (\ref{Phi1s}) is very time-consuming. An approximation can be given comparing with the solution for the harmonic oscillator \cite{wir}
\begin{eqnarray}
\label{rho2s}
&&\varrho_{2s^4}^{\rm cm, HO}(a, R)=|\psi^{\rm HO}_{2s^4}( R)|^2=\left(\frac{a}{\pi}\right)^{3/2} e^{-4a R^2} 
\nonumber \\&& \times
\frac{1}{10616832} (24695649+14905152\, a R^2+354818304\, a^2R^4 
\nonumber \\&&
-876834816\, a^3R^6
+1503289344\, a^4R^8-1261699072\, a^5R^{10} 
\nonumber\\ &&
+613416960\, a^6R^{12}-150994944\, a^7 R^{14}+16777216\, a^8R^{16}).
\nonumber\\ &&
{}
\end{eqnarray}
The parameter $a''=0.287038$ fm can be chosen to reproduce the value at $R=0$ (three-fold integral). The scaling $R''=f_{\rm scal}(R) + 0.174 \,(e^{R/2.924}-1)$ fulfills normalization and improves the asymptotic behavior for large $R$, so that $\varrho_{2s^4}^{\rm cm}(R)\approx \varrho_{2s^4}^{\rm cm, HO}(a'', R")$. 
A plot of $(4 \pi R^2)^{1/2} \psi_{2s^4}( R)$ is shown in Fig. \ref{fig:3}. The normalization $\int_0^\infty 4 \pi R^2 \psi^2_{2s^4}( R) dR =1$ holds.

\begin{figure}
\resizebox{0.5\textwidth}{!}{%
  \includegraphics{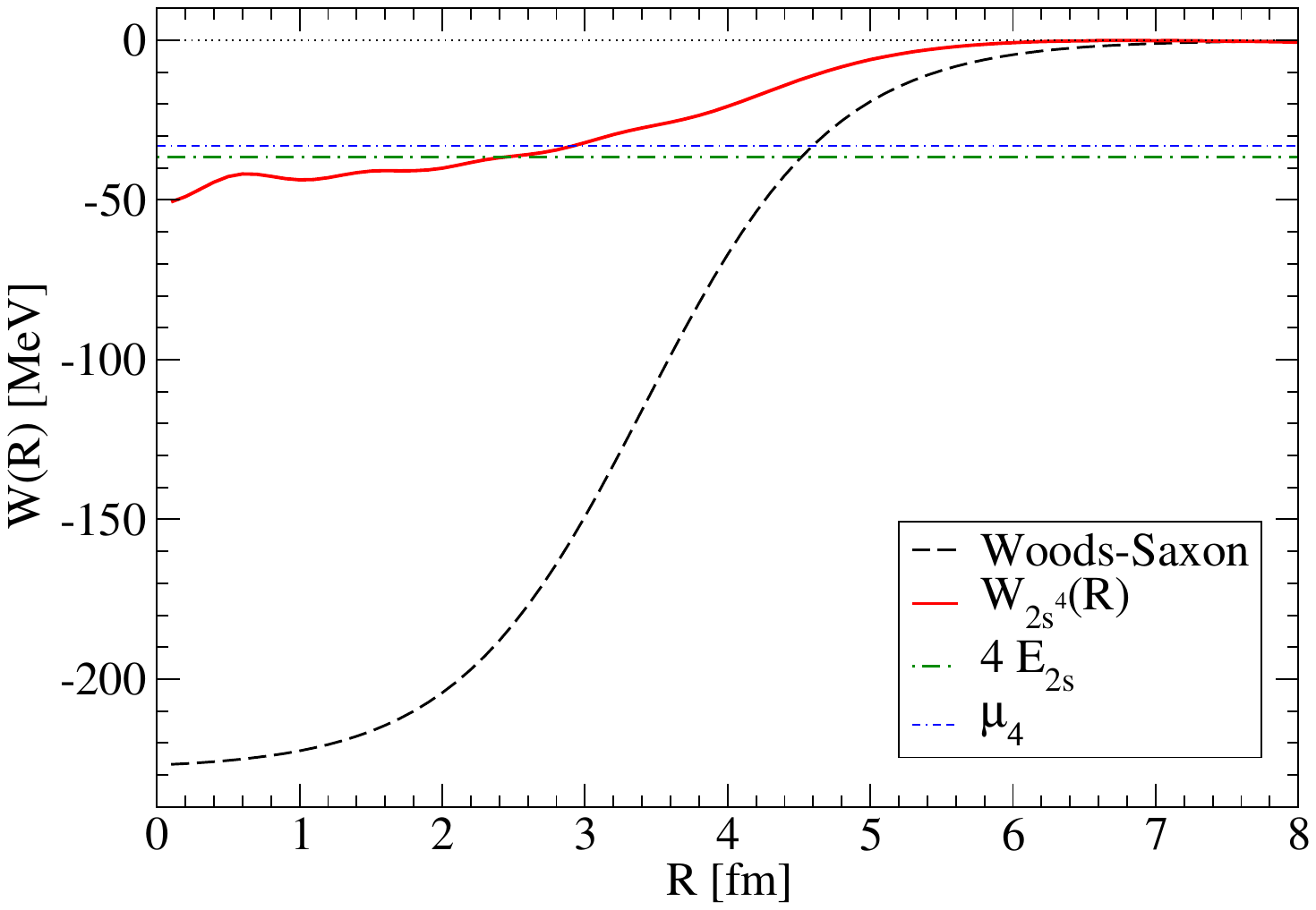}
}
  \caption{The c.m. potential $W_{2s^4}( R)$, Eq. (\ref{Sglcom1}), compared with the Woods-Saxon potential of the quartet.}
\label{Fig:W}
\end{figure}

We reconstruct the effective potential from the wave function  
$\psi_{2s^4}( R)=(\varrho_{2s^4}^{\rm cm}( R) )^{1/2}$  \cite{wir}.
If we restrict us to $s$ states ($l=0$) and introduce $u_{2s^4}( R)= (4 \pi)^{1/2} R \psi_{2s^4}( R)$, 
we have
\begin{equation}
\label{Sglcom1}
W_{2s^4}( R)-E_{2s^4}=\frac{\hbar^2}{8m} \frac{1}{u_{2s^4}( R)} \frac{d^2}{dR^2} u_{2s^4}( R).
\end{equation}
The result is shown in Fig. \ref{Fig:W}.

We conclude from this: The effective c.m. potential $W( R)$ remains  almost constant within the core 
as expected from the Thomas-Fermi model. The value $E_{2s^4} = -36.65$ MeV 
is near to the estimate $\mu_4=-33$ MeV from the Thomas-Fermi rule.
It is slightly increasing near the surface, possibly because the quartet is not localized at a point,
but smeared out, so that it "feels" the weakening of the potential near the surface. 
Another reason could be the gradient terms in Eq. (\ref{9}), which are neglected here.
A similar behavior was also observed for the harmonic oscillator potential in \cite{wir}.
In contrast to the harmonic oscillator, where the effective potential increases with $R$, 
the behavior near the surface is now more realistic.
The weakening of the Thomas-Fermi rule has been shown in Refs. \cite{Po14,Xu16,Xu17,wir}.

\subsection{Intrinsic interaction and Pauli blocking}

We have introduced an effective c.m. potential $W( R)$, which describes the influence of the environment 
(here the core nucleus) on the c.m. motion of the quartet in mean-field approximation. 
Specifically, we have simulated a quartet of 4 uncorrelated nucleons in $2s$ states moving under the influence of the core nucleus $^{16}$O. 
The corresponding potential $W_{2s^4}( R)$ shows approximately the constancy of the chemical potential required within the Thomas-Fermi model.

\begin{figure}
\resizebox{0.5\textwidth}{!}{%
  \includegraphics{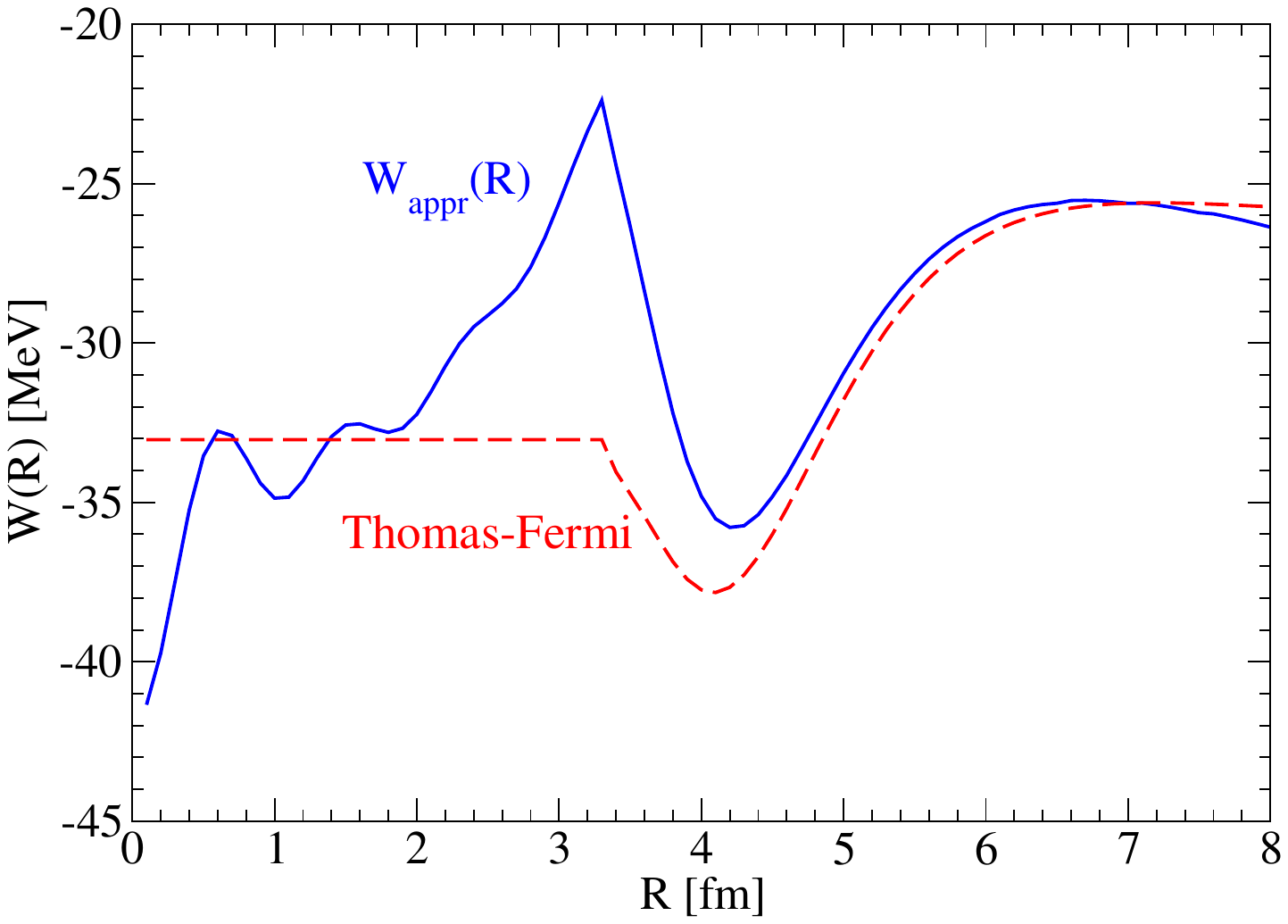}
}
  \caption{Quartet c.m. potentials $W(R )$. The Thomas-Fermi approximation $W^{\rm TF}(R )$ 
is compared with the calculation $W_{\rm appr}(R )$ using harmonic oscillator shell model states. Note the peak at $r_{\rm crit}=3.302$ fm.}
\label{Fig:pocket}
\end{figure}

To describe the formation of an $\alpha$-like cluster, we need to consider the interaction within the quartet.
To estimate the intrinsic interaction of the quartet, we add for $R > r_{\rm crit}$ the energy shift $W^{\rm intr}({\bf R})$, Eq. (\ref{WeffR}), which
describes the formation of the cluster and the dissolution due to Pauli blocking,
see fig. \ref{Fig:pocket}. 
The Coulomb potential is added, and the free effective potential of the shell model 
$W_{2s^4}( R)$ is used instead of $W^{\rm ext}$. 
A harmonic oscillator base was essentially used here \cite{Ro18}.
We denote this approximation for the potential for the c.m. motion as $W_{\rm appr}(R )$.

Obviously, this c.m. potential $W_{\rm appr}(R )$ is only a rough approximation.
In particular, the sharp peak due to the sudden switching off of the intrinsic interaction at $r_{\rm crit}=3.302$ fm does not seem realistic.
A similar peak at $r_{\rm crit}$ was also obtained for the heavy isotopes \cite{Xu16,Xu17}, but it was less pronounced than for the light isotope $^{20}$Ne.

The behavior for large $R$ is correctly reproduced, the asymptote $\lim_{R \to \infty} W(R)=-28.3$ MeV is the binding energy of the $\alpha$ particle, and the Coulomb repulsion is well represented. The attractive $N-N$ interaction is also visible, as in other approaches using an optical potential, see App. \ref{app:1}. As the density of the core increases, the binding energy of the $\alpha$ cluster is weakened due to Pauli blocking, and a pocket is formed.
The behavior for small $R \le 2$ fm is also well reproduced. The fluctuations around the Thomas-Fermi value are due to the shell structure. 

An improvement of the effective quartet potential is particularly necessary in the vicinity of the critical density. Instead of a sharp switchover, in which all correlations above the critical density are omitted, these decrease continuously. Quartet correlations are also present for 
$R \le r_{\rm crit}$. They can provide a contribution as resonances in the continuum, which decreases steadily with increasing density. Furthermore, Pauli-blocking is calculated for uncorrelated nucleons in the environment, which is expressed in the use of the Fermi function. Correlations in the surrounding matter would also reduce the Pauli blocking. Taking into account the c.m. movement of the $\alpha$-cluster, the Pauli blocking is also reduced.

Furthermore, we are dealing with an inhomogeneous system, so that gradient terms can become important. 
As an extended system, the $\alpha$-like cluster is determined not only by the properties of the surrounding matter at the position of the center of mass, 
but by the properties within the extension of the cluster. Finally, the Pauli principle is a non-local effect, which is treated as local only after some approximations.
We have collected several arguments which show that the effect of Pauli blocking should be treated as a continuous function of density.
This can help to reduce the peak at the critical radius.

\subsection{Shell-model calculations}

\begin{figure}
\resizebox{0.45\textwidth}{!}{%
  \includegraphics{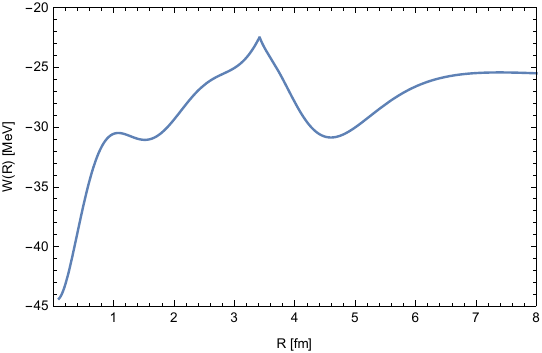}
}
  \caption{Quartet c.m. potential $W(R )$ for $^{20}$Ne using shell model states.}
\label{Fig:QWFA20Ne}
\end{figure}

First results to use shell-model calculations for $^{20}$Ne to perform calculations within the QWFA have been presented in  Refs. \cite{Yang21,Bai19}. We use the widely-used Woods-Saxon potential
\begin{equation}
\label{vws}
V_{\rm WS}\left ( r \right ) = \frac{V_{0}}{1+\textrm{exp}(\frac{r-R_0}{a})},
\end{equation}
together with the spin-orbit coupling interaction
\begin{equation}
\label{vso}
V_{\rm so}\left ( r \right ) = \frac{1}{2\mu^2 r}\left ( \frac{\partial }{\partial r} \frac{\lambda V_{0}}{1+\textrm{exp}(\frac{r-R_{\rm so}}{a_{\rm so}})}\right ) \bf l \cdot \bf s
\end{equation}
to determine the shell model wave functions of quartet nucleons in $^{20}$Ne. 
The strength of the Woods-Saxon potential is parameterized as 
\begin{equation}
V_{0}=-46\left [ 1\pm 0.97\left( \frac{N-Z}{A} \right) \right ]
\end{equation}
(``$+$" for protons and ``$-$" for neutrons). 
The parameter $R_0$ is $1.43\,A^{1/3}$ fm for both protons and neutrons while the parameter $R_{\rm so}$ is $1.37\,A^{1/3}$ fm. 
The diffusivity parameter $a$ and $a_{\rm so}$ are chosen to be the same value 0.7 fm. $\mu$ is the reduced mass of the $\alpha$-core system and the normalization factor of the $ls$ coupling strength $\lambda$ is 37.5 for neutrons and 31 for protons, respectively. 
The Coulomb potential we adopt is
\begin{eqnarray}
\label{vcoul}
V_C(r)&=&(Z-1)e^2  (3R_{\rm Coul}^2-r^2)/2R_{\rm Coul}^3,\quad  r\le  R_{\rm Coul}, \nonumber \\ &=& (Z-1)e^2 /r,\quad r> R_{\rm Coul}.
\end{eqnarray}
with the radius $R_{\rm Coul}=1.25\,A^{1/3}$ fm. 
The effective c.m. potential constructed from the shell model quartet state for $^{20}$Ne is shown in Fig. \ref{Fig:QWFA20Ne}. 

A general discussion of the Pauli blocking term is necessary to avoid the peak in Figs. \ref{Fig:pocket} and \ref{Fig:QWFA20Ne}.
Various approximations were made when calculating the effective potential. 
We mention the neglect of the gradient terms and the non-local property of the potential $W({\bf R},{\bf R}')$, Eq. (\ref{eq:10}), in particular due to the Pauli blocking term.
We emphasize that Eq. (\ref{WPauli}) was derived for $\alpha$-particles in an uncorrelated medium. At zero temperature, the medium can be strongly correlated and form $\alpha$ matter. 
A correlated medium was considered in Ref. \cite{Tak04}, and the merging with the continuum was observed at a slightly higher critical density. 
If we use this calculation to construct the Pauli blocking shift, this could possibly lead to a smoother transition and reduce the peak.

\begin{figure}
\resizebox{0.5\textwidth}{!}{%
  \includegraphics{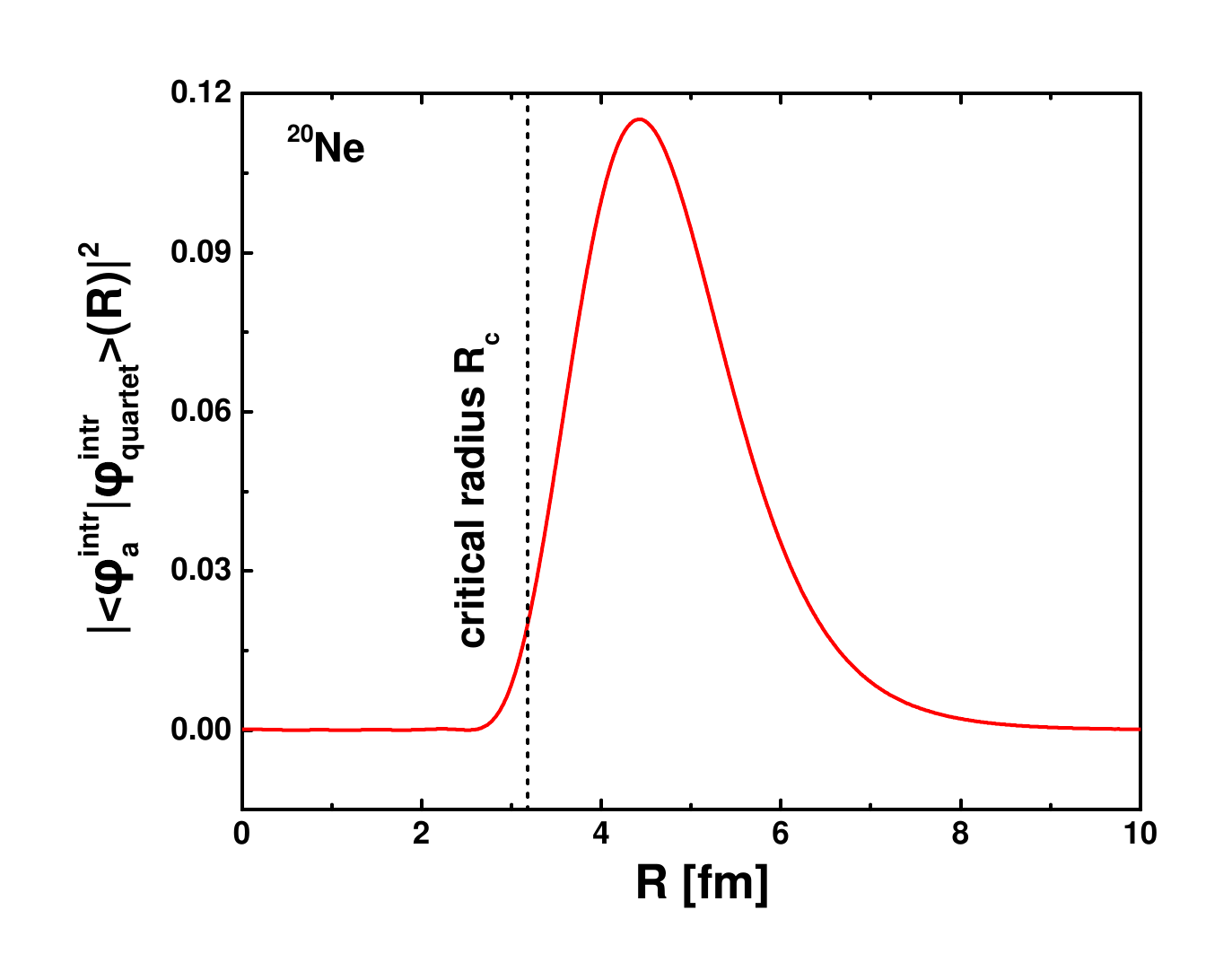}
}
  \caption{The overlap between the intrinsic wave functions of the quartet and the $\alpha$-particle as a function of c.o.m. coordinate $R$ for the $\alpha$+doubly magic core system $^{20}$Ne.}
\label{Fig:Overlap20Ne}
\end{figure}

For $^{20}$Ne, the probability to find the $\alpha$-particle in the localized shell model states can be defined as
\begin{eqnarray}
{\cal F}_\alpha=\int dR\,4\pi R^2 \rho_{\rm quartet}^{\rm c.m.}(R)\left| \left \langle\varphi_{\alpha}^{\rm intr} | \varphi_{\rm quartet}^{\rm intr}\right \rangle(R)\right|^2,
\end{eqnarray}
where $\left \langle\varphi_{\alpha}^{\rm intr} | \varphi_{\rm quartet}^{\rm intr}\right \rangle(R)$ is the overlap between the intrinsic wave functions of a quartet $\varphi_{\rm quartet}^{\rm intr}$ and a free $\alpha$-particle as a function of c.m. variable $R$. 
The density at the c.m. position $\bf R$ is $\rho_{\rm quartet}^{\rm c.m.}(R)=\mid{\Psi_{\rm quartet}^{\rm c.m.}(\bf R)}\mid^2$. 
As expected, the probability ${\cal F}_\alpha =2.004\times 10^{-3}$ is quite small for $^{20}$Ne as the wave function of the quartet is approximated by a product of shell model states. 
However, the probability ${\cal F}_\alpha$ is significantly enhanced for the $\alpha$ + doubly magic core system $^{20}$Ne as  compared to those of their neighboring isotopes
We show in Fig. \ref{Fig:Overlap20Ne} the overlap between the wave functions of the quartet and the $\alpha$-particle as a function of c.m. coordinate $R$ for $^{20}$Ne. 
It is clearly demonstrated that there exists a peak in the region beyond the critical radius (i.e. the surface region of the core). 
Inside the core, the probability to find the $\alpha$-like state is quite low for $^{20}$Ne.

\section{Comparison with the THSR model and other approaches}
\label{sec:THSR}

\subsection{Calculations for $^{20}$Ne}

\begin{figure}
\resizebox{0.47\textwidth}{!}{%
  \includegraphics{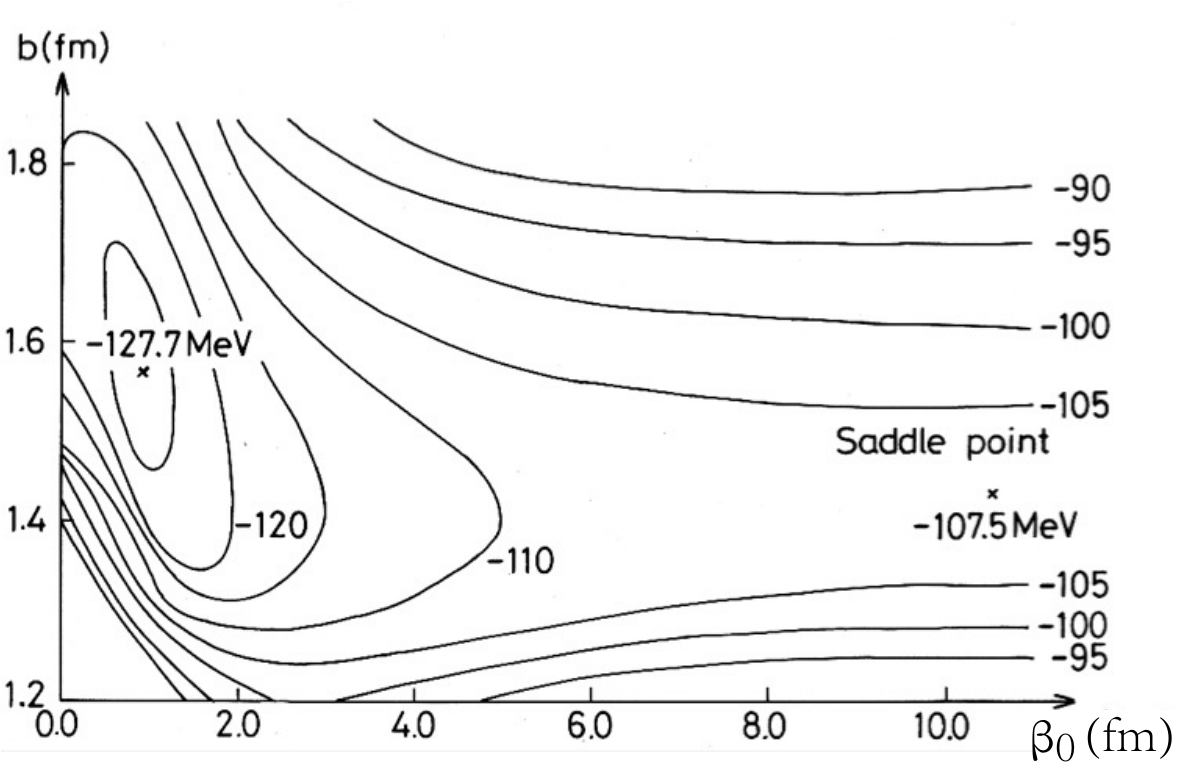}
} 
\caption{ Variational calculations for the energy of $^{16}$O with respect to the harmonical osciallator parameter $b$ and size parameter $\beta_0$ using the THSR wave function~\cite{THSR}.}
\label{fig:9}    
\end{figure}
 
The THSR ansatz adeptly describes the low-density regime of \(\alpha\) matter as well as the shell model states, particularly when the c.m. wave function coincides with the intrinsic wave function. Notably, when four \(\alpha\) clusters merge into a \(^{16}\mathrm{O}\)-like configuration, the antisymmetrization process gives rise to nucleonic $s$ and $p$ orbitals, especially as the inter-cluster distance approaches zero. Deviations in the Gaussian width parameters signal the presence of correlations. 
The $N\alpha$ THSR wave function~\cite{THSR} can be written as, 
\begin{equation}
\Phi_{n\alpha}^{\rm THSR}\!\!\propto {\cal A}\ \Big\{ \prod_{i=1}^n\exp \Big[-\frac{2 (\vec{X}_{i}-\vec{X}_{G})^2}{b^2+2\beta_0^2} \Big] \phi(\alpha_i) \Big\}, 
\label{eq:int_thsr}
\end{equation}
where $\vec{X}_{i}$ and $\vec{X}_{G}$ are the c.m. coordinate of the $\alpha$ cluster and the total c.m. coordinate of $N\alpha$ cluster, respectively. 
Figure \ref{fig:9} presents a THSR calculation for \(^{16}\mathrm{O}\), where the parameter $\beta_0$ reflects deviations from shell model behavior. 
The observed energy minimum at a finite $\beta_0$ (size parameter of the THSR wave function) signifies the existence of \(\alpha\)-like correlations even in the ground state. 

The uncorrelated mean-field approximation, often invoked to compute Pauli blocking effects, may not be universally valid. In particular, \(\alpha\) matter exemplifies a scenario where the medium undergoes a transformation into a correlated state. Analogous reconfigurations are evident in pairing phenomena at temperatures descending below the critical value. The Tohsaki-Horiuchi-Schuck-R\"opke (THSR) formalism was conceived to elucidate \(\alpha\) clustering within such tenuous nuclear environments, exemplified by the Hoyle state of \(^{12}\mathrm{C}\). Here, the environment of an \(\alpha\) cluster is composed of other \(\alpha\) clusters, leading to a pronouncedly clustered structure. This method has been successfully employed to investigate various $4n$ nuclei, 
including \(^{20}\mathrm{Ne}\).

\begin{figure}
\resizebox{0.45\textwidth}{!}{%
  \includegraphics{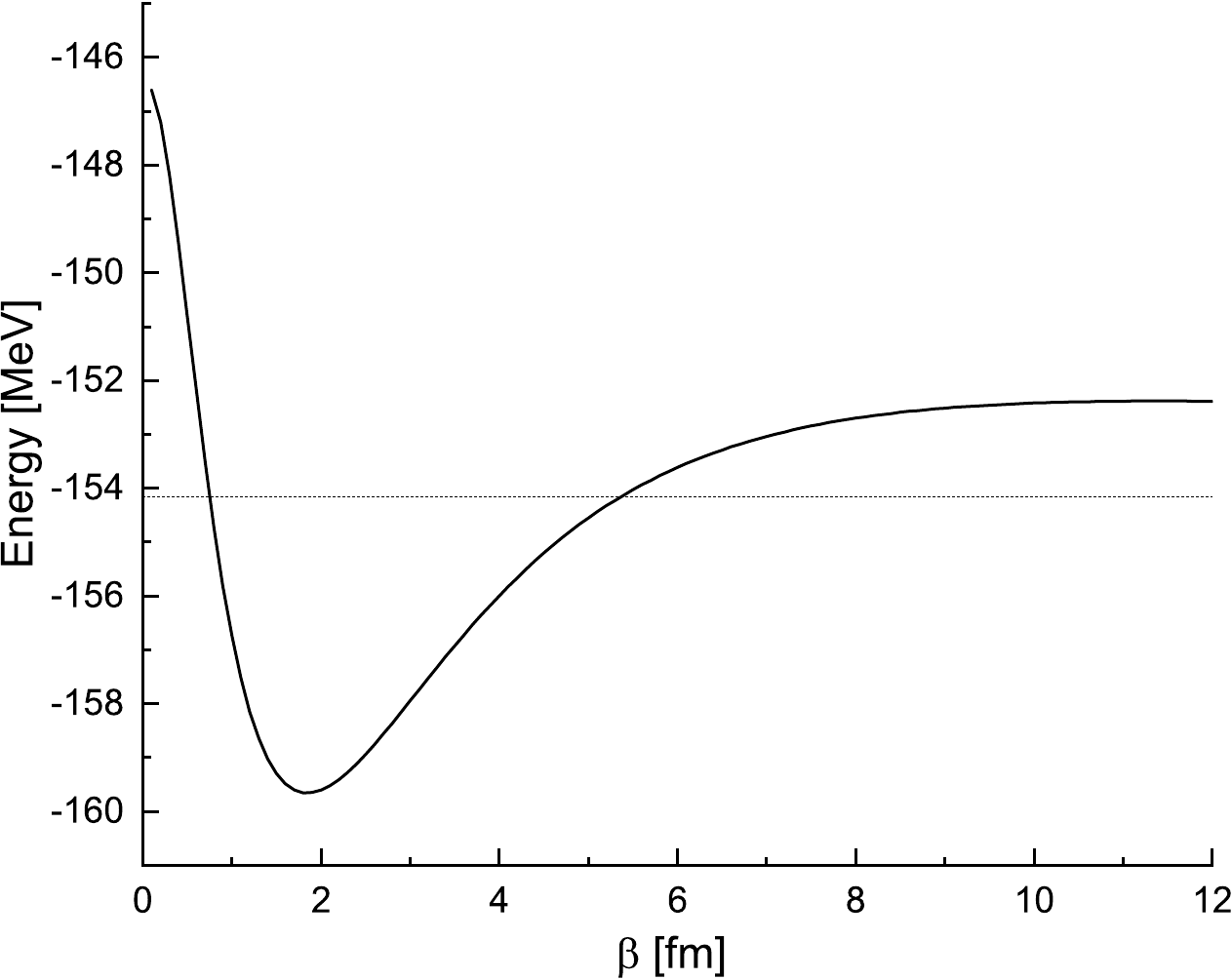}
}
\caption{Energy curve of  $^{20}$Ne with the increase of the size parameter $\beta$ using the intrinsic THSR wave function. The asymptotic -154.16 MeV for the binding energy of separated $^{16}\mathrm{O}$ and \(\alpha\) clusters is also shown. 
}
\label{fig:9a}    
\end{figure}

The microscopic THSR wave function for the nucleus \(^{20}\mathrm{Ne}\) can be written as 
\begin{eqnarray}
\label{THSR}
{\widehat \Phi}_{{\rm THSR}}(\beta) = {\cal A}[\exp(-\frac{8r^2}{5(b^2+2\beta^2)}\phi(\alpha)\phi({^{16}{\rm O}})],
\end{eqnarray}
where ${\bf r}={\bf X}_{1}-{\bf X}_{2}$. ${\bf X}_{1}$ and ${\bf X}_{2}$ represent the center-of-mass coordinates of the $\alpha$ cluster and the $^{16}\mathrm{O}$ cluster, respectively. It should noted that the  $^{16}\mathrm{O}$ cluster is described as the shell model wave function.

Fig. \ref{fig:9a} shows the energy curve with the increase of the size parameter $\beta$.
This can be transformed to the energy curve as a function of the inter-cluster distance. 
The extracted effective $\alpha$-O potential would be of interest. 
It should be noted, however, that the inter-cluster distance between clusters cannot be precisely defined, especially when clusters are in close proximity, owing to the effects of antisymmetrization.

It is not directly possible to define the inter-cluster distance $D$ in THSR approach. According to Matsuse  \cite{Mat75} one can introduce the distance $D$ according the relation for the rms radii
\begin{equation}
    20 \langle r^2 \rangle_{\rm Ne}=16 \langle r^2 \rangle_{\rm O}+4\langle r^2 \rangle_{\alpha}+\frac{16}{5}\langle D^2 \rangle
\end{equation}
so that 
\begin{equation}
    \langle D^2 \rangle = \frac{25}{4} \langle r^2 \rangle_{\rm Ne}-\frac{195}{16} b^2.
\end{equation}
follows. We used this quantity $D$ for the distance $r$ in Fig. \ref{fig:Coul}.

\begin{figure}
\resizebox{0.4\textwidth}{!}{%
\includegraphics{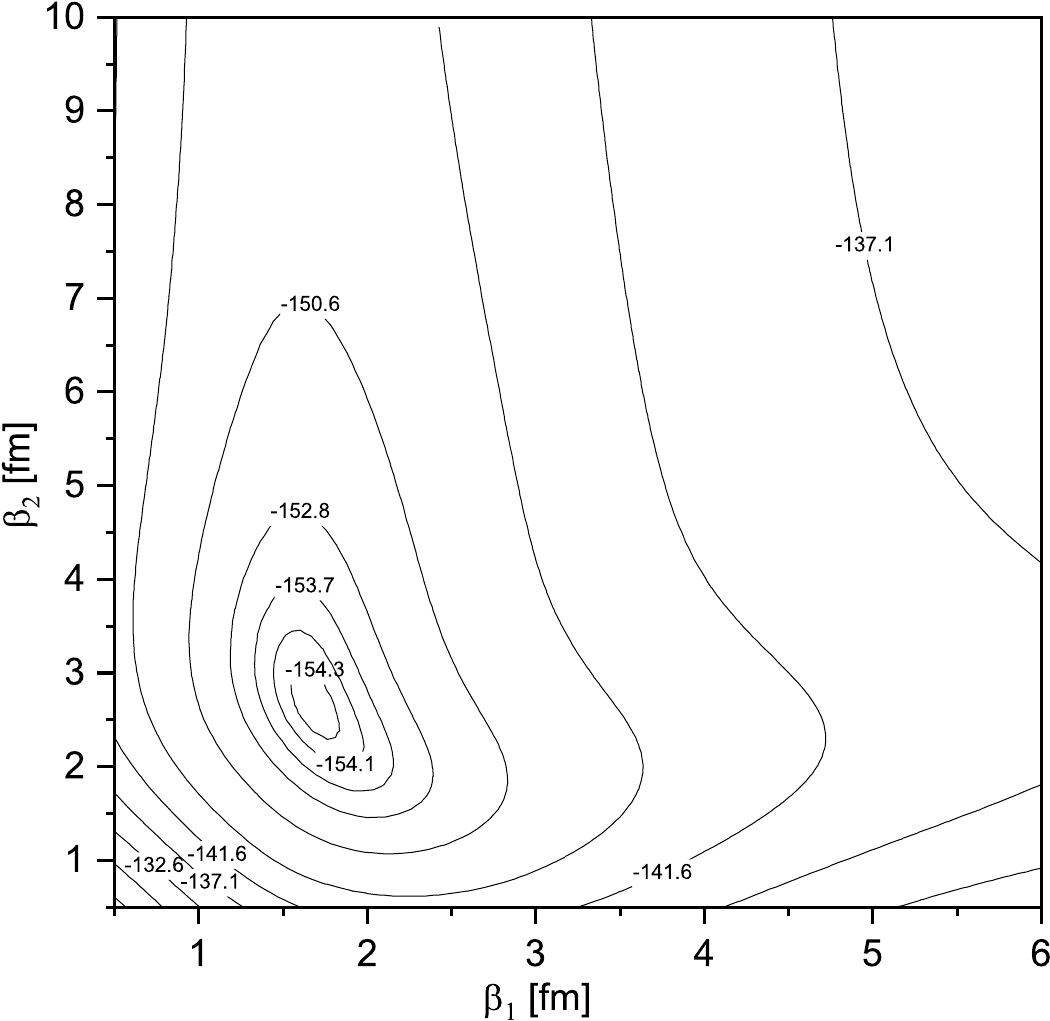}
}
\caption{Contour plot for the ground state of $^{20}\mathrm{Ne}$ in the spherical $\beta_1$ and $\beta_2$ parameter space. }
\label{fig:Bo}    
\end{figure}

\begin{figure}
\resizebox{0.5\textwidth}{!}{%
  \includegraphics{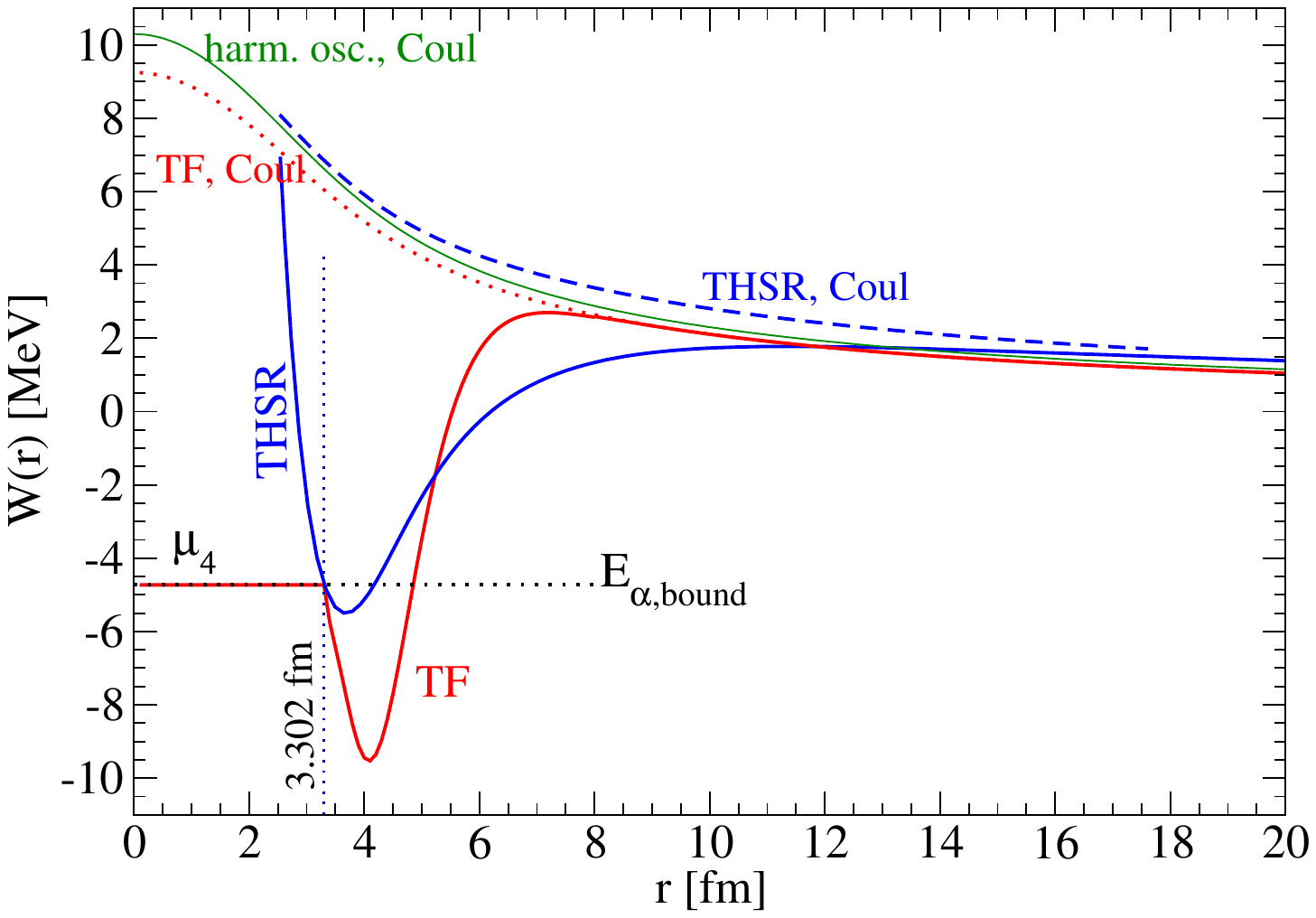}
}
\caption{$^{16}$O - $\alpha$ effective interaction potential as function of the center-of-mass distance $R$.
The THSR calculations (blue full line) are compared with the Thomas-Fermi approximation of the quartetting wave function (red full line). 
The total potential (TF, THSR) is shown as well as the Coulomb contribution (dashed lines). In addition, the Coulomb interaction for the
harmonic oscillator density of the O-core is also shown.}
\label{fig:Coul}    
\end{figure}

Very recently, the $5\alpha$ clustering structure of $^{20}\mathrm{Ne}$ was scrutinized by Bo {\it et al.}  \cite{Bo23} utilizing the THSR framework, which adopts the container model. In this model, the intrinsic $\alpha$ cluster width parameter $b$ is complemented by two additional parameters: $\beta_1$ (denoting the width of the $^{16}\mathrm{O}$ core nucleus) and $\beta_{2}$ (representing the center-of-mass motion of the residual $\alpha$ cluster). 
As illustrated in Fig. \ref{fig:Bo}, the energy minimum is observed at $\beta_1 = 1.5$ fm and $\beta_2 = 3.0$ fm, corresponding to an energy of approximately $-155.3$ MeV. The GCM calculations yield an energy of $-156.4$ MeV. Additionally, the calculated rms radius is $2.96$ fm. 
A notable aspect of the THSR wave function is its inclusion of the shell model limit, thereby ensuring an accurate representation of the ground state of the $^{16}\mathrm{O}$ core nucleus. The orthogonality between the additional fifth $\alpha$ particle and the core states is rigorously preserved. Theoretical calculations yield favorable comparisons with empirical data for both the binding energy and the root-mean-square (rms) radius of the ground state. The disparity between the values of $\beta_1$ and $\beta_2$ suggests the presence of an $\alpha$ particle atop the doubly-magic $^{16}\mathrm{O}$ core.
These results from the $5\alpha$ calculations set the stage for future studies to develop a more realistic $^{16}\mathrm{O}$-$\alpha$ effective interaction.

%

%

\subsection{$\alpha$ matter}

The equilibrium composition of homogeneous nuclear matter at low densities and temperatures is a complex problem, since below the saturation density of symmetric matter $\rho_{\rm sat}=0.15$ fm$^{-3}$ a thermodynamic instability occurs and clusters are formed. The highest binding energy per nucleon is found for the nucleus $^{56}$Fe. Here we only consider the formation of $\alpha$-clusters from the nucleons. 

At a fixed baryon density, the mass fraction of the $\alpha$ clusters increases with decreasing temperature. 
At a critical temperature, a quantum condensate can be formed. 
In analogy to pairing, the $\alpha$-like quartets are the bosonic components of the condensate. 
However, they are modified by the medium \cite{Fun08}.
As known from pairing, where the Bogoliubov transformation allows to describe the nuclear matter below the critical temperature, below the critical temperature for quartetting we have to consider a correlated medium, the so-called $\alpha$ matter.

In analogy to the THSR approach for low-density nuclei such as $^8$Be or the Hoyle state of $^{12}$C, calculations for periodic $\alpha$-like structures were performed in \cite{Tak04}. Orthonormal Bloch states were introduced so that Pauli blocking by nucleons bound in $\alpha$-clusters is strictly realized. One problem is the separation of the c.m. contribution to the kinetic energy, which is solved by a simple ansatz  based on the energy gap at zero momentum. As a result, in Ref. \cite{Tak04} it was shown that the bound state merges with the continuum at about $0.2 \rho_{\rm sat}$.

\begin{figure}
\resizebox{0.5\textwidth}{!}{%
  \includegraphics{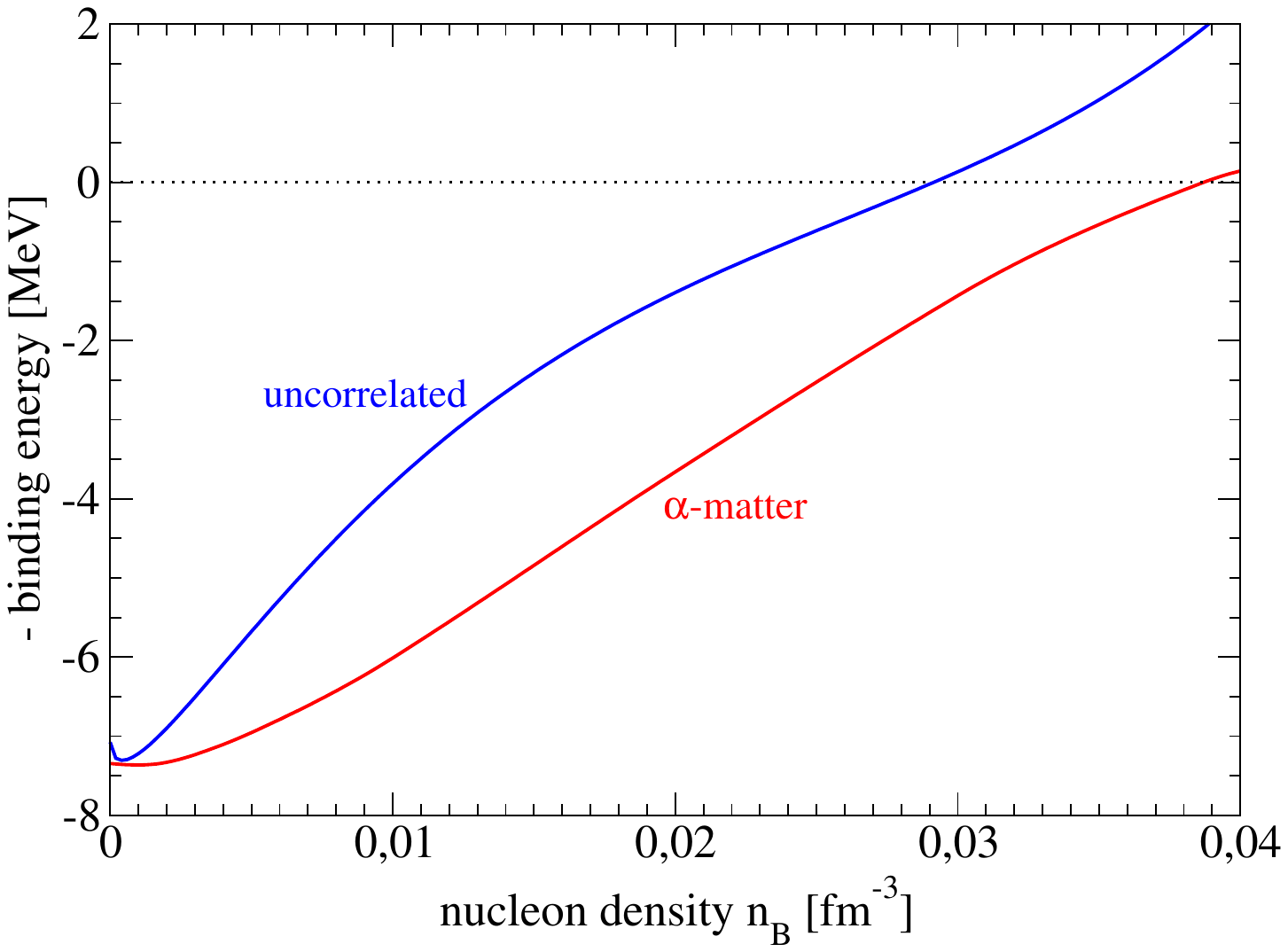}
}
\caption{Shift of the binding energy per nucleon for an $\alpha$-cluster as function of the nucleon density $n_B$.
The difference of the energy per nucleon in $\alpha$-matter and in momentum eigenstates (red) is compared with the shift (blue) in uncorrelated matter, Eq. (\ref{eq:Paul}).}
\label{fig:paul}    
\end{figure}

We have also performed exploratory calculations with a separable potential adapted to reproduce the free-$\alpha$ properties mass and rms radius, see Appendix \ref{app:shifts}. 
The difference between the energy per nucleon in the uncorrelated free-nucleon state and the $\alpha$-matter state is shown in Fig. \ref{fig:paul}.
A value $\rho_{\rm Mott}=0.04$ fm$^{-3}$ was found for the dissolution of the bound state. 

For comparison, in Fig. \ref{fig:paul} also shown is the shift of the binding energy for uncorrelated matter where the surrounding nuleons occupy free single-particle states,
\begin{equation}
\label{eq:Paul}
    E_{\rm bound}^{\rm uncorr}(n_B) =-7.07\, {\rm MeV}+W^{\rm Pauli}(n_B)-E_F(n_B).
\end{equation}
Compared to the Pauli blocking by free nucleons considered in Eq. (\ref{WPauli}), the blocking in $\alpha$-matter is smaller because the distribution in momentum space is spreed out, and the blocking is less efficient. As a result, the critical density where bound states are dissolved, comes out to be larger if cluster formation in the surrounding is taken into account.
We expect that this modification makes the  peak in the figures \ref{Fig:pocket} and \ref{Fig:QWFA20Ne} smoother. Further investigations are necessary to find a better treatment of the dissolution of clusters due to Pauli blocking.

Another approach to show that nuclear matter dissolves into clusters at low density was presented in Ref. \cite{Ebran20}. Restricted Hartree-Fock calculations were performed that allow the formation of separate cluster structures. 
Even with this approach, the strict separation of the kinetic energy of the c.m. remains open.
An unresolved question is whether the disappearance of the cluster structures and the appearance of a homogeneous phase is a first-order transition.

\subsection{Other approaches to $\alpha$-clustering in nuclei}

Based on a local density approach with composition and energy shifts derived from \cite{R1,R11}, Typel \cite{T14} has considered the formation of $\alpha$-particle correlations at the surface of heavy nuclei to study the neutron skin thickness of heavy nuclei and the slope of the nuclear symmetry energy. 
The $\alpha$ particle density was considered as a function of radius for the tin isotopes $^{108}$Sn to $^{132}$Sn, and it was shown that as the neutron density at the skin increases, the abundance of $\alpha$-particles is suppressed as a result of Pauli blocking. The experimental evidence for the $\alpha$ cluster formation in the surface of neutron-rich tin isotopes was given using quasi-free $\alpha$ cluster-knockout reactions \cite{T20,T21}. Note that the occurrence of $\alpha$-cluster at the surface of $^{48}$Ca and $^{208}$Pb and and its impact on the extraction of symmetry energy from skin thickness is also investigated by using QWFA \cite{Yang23}. Strong closed shell structure effects and complex derivative terms of the intrinsic wave function are properly taken into account in QWFA \cite{Yang23}.

The question of $\alpha$ formation in the ground state of heavy nuclei has been investigated using the AMD approach  \cite{AMD,AMD04} in several recent publications. 
The AMD approach also describes the suppression of clusters using the Pauli blocking effect. The manifestation of clustering at the surface region where the density is low has induced many investigations on $\alpha$-break-up reactions.
We do not give a comprehensive account of various investigations of specific isotopes \cite{Freer18,Yos19,Chi17,Tan21,Yos22,Qi10,Nak23,Kimura22,PG11} in this paper. 
We would like to emphasize that the approach of the quartet wave function presented here is also of interest for these examples.

\section{Conclusions}
\label{sec:Con}

We investigated the c.m. motion of an $\alpha$-like quartet, which moves under the influence of a 
core nucleus, here the $^{16}$O nucleus. In local density approximation, 
an effective potential $W( R)$ for the quartet c.m. motion 
is obtained, which shows a pocket structure near the surface of the nucleus. 
This is important for the preformation of $\alpha$ particles near the surface. 
A new aspect is the behavior of $W( R)$ inside the core nucleus, 
i.e. for $R \le r_{\rm crit}$, where the quartet bound state is dissolved due to Pauli blocking.
In contrast to earlier studies, which assume an increase in the
effective $\alpha- ^{16}$O-potential with decreasing $R$, in a Thomas-Fermi approach  
$W^{\rm TF}(R)=\mu_4$ remains constant in this range $R \le r_{\rm crit}$ \cite{Po14,Xu16,Xu17,wir}.
In the present work, we also show for
the shell model approach that the effective potential $W(R )$ remains almost constant in the core nucleus. 
 The reason for this is the exchange interaction or Pauli blocking 
between the quartet nucleons and the core nucleus.

For large distances, the empirically determined M3Y potential used for $W(R )$
agrees with the optical potentials derived from scattering experiments. Near the surface of the nucleus 
the Pauli blocking becomes relevant. A pocket that is formed for the effective potential $W^{\rm TF}(R)$  
is also retained after the introduction of single-particle shell model states for the core nucleus. 
However, the local density approximation for the Pauli blocking should be improved,
and it is expected that sharp peak structures observed for $W(R )$ in shell model calculations will be smeared out. 

Of interest is the comparison with the THSR approach \cite{THSR,Toh17}, which treats the quartets 
 self-consistently. If a mean-field description for the surrounding medium based on 
uncorrelated single-particle states is no longer possible, correlations in the medium, 
especially quartetting, should be taken into account. The full antisymmetrization 
of the many-body wave function is a great challenge. The THSR
approach offers us such a self-consistent, antisymmetrized treatment of quartetting of all nucleons.
A variational principle with Gaussian wave functions was used, and nuclei with $A \le 20$
were treated in this way. Interesting results were obtained for $^{20}$Ne \cite{Bo,Bo2,Bo3} 
considering the full antisymmetrization of the $\alpha$- and $^{16}$O-wave functions . 
We have tried to find appropriate observables in the THSR approach to derive an effective potential $W( R)$ and a wave function $\psi( R)$ for the quartet c.m. motion, 
to compare them with the quartetting wave function approach.

Our general vision is to treat quartetting in the nuclear matter self-consistently, as is the case for pairing.
The approaches described in this paper provide only partial answers to this project. 
The THSR approach comes closest to this goal, but it contains some restrictions, so that it is not generally applicable.
Although the quartet wave function approach is generally applicable, it contains several approximations that still need to be improved.
One main problem is the treatment of Pauli blocking. 
The local approximation with a cut-off of $\alpha$-like clusters at the critical density needs to be improved in future work.

\subsection*{Acknowledgement}
We would like to dedicate this work to the memory of our esteemed colleague and friend, Peter Schuck, with whom we have had the privilege of collaborating for many years. Peter's broad interests and profound insights in nuclear physics have been an inspiration to us all. We are deeply grateful for his companionship and contributions throughout the years. He will be dearly missed. His spirit and dedication to the pursuit of knowledge continue to guide us, and in his memory, we commit to advancing the work he so passionately embraced.

C. Xu is supported by the National Natural Science Foundation of China (Grant No. 12275129).
This work was supported in
part by the National Natural Science Foundation of China under contract Nos. 12175042,12147101.
Zhongzhou REN thanks the support of National Natural Science Foundation of China
with grant number 12035011.
The work of G.R. received support via a joint stipend from the Alexander von Humboldt Foundation
and the Foundation for Polish Science.

\appendix{

\section{Optical model description and double-folding potential}
\label{app:1}

We discuss the effective c.m. potential $W^{\rm TF}( R)$ and compare with other approaches, see also \cite{wir}. 
In particular, we check
whether the choice (\ref{cd}) for the double-folding potential parameters $c,\,\,d$ are realistic.
Several approaches to the optical potential are shown in Fig. \ref{fig:W}.

The elastic scattering of $\alpha$ particles on the $^{16}$O nucleus was investigated, 
and the corresponding optical potentials were inferred. There is a large uncertainty 
for small values of $R$.
A first expression for the real part of the optical potential is  \cite{Fadden66}
\begin{equation}
\label{optpot}
-\frac{V_0}{1+e^{(r-r_0A^{1/3})/a}}
\end{equation}
with $V_0=43.9$ MeV, $r_0=1.912$ fm and $a=0.451$ fm.
Improvements were madein Ref. \cite{Fukui16} considering the $^{16}$O ($^6$Li,d) $^{20}$Ne transfer reaction, where the  model potential (\ref{optpot})
with $r_0=1.25$ fm and $a=0.76$ fm was used, $V_0$ was adjusted to reproduce the value 4.73 MeV of the binding energy. 
Kumar and Kailas \cite{Kumar} give the parameter values $V_0=142.5$ MeV, 
$r_0=1.18$ fm and $a_0=0.76$ fm.

Another approach \cite{Michel} was compared with experiments \cite{Oertzen,Oertzen1}. They used the expression
\begin{equation}
-V_0\frac{1+\alpha e^{-(r/\rho)^2}}{[1+e^{(r-R_R)/(2 a_R)}]^2}
\end{equation}
with $V_0=38$ MeV, $\rho = 4.5$ fm, $R_R=4.3$ fm, $a_R=0.6$ fm, and the energy-dependent $\alpha = 3.625$.
More recently, in Ref. \cite{Ohkubo}  a density dependent effective M3Y 
interaction (DDM3Y) was used, and a double-folding potential was derived (Fig. 3 in \cite{Ohkubo}) which ranges at $R=0$ to -110 MeV.

\begin{figure}
\resizebox{0.5\textwidth}{!}{%
  \includegraphics{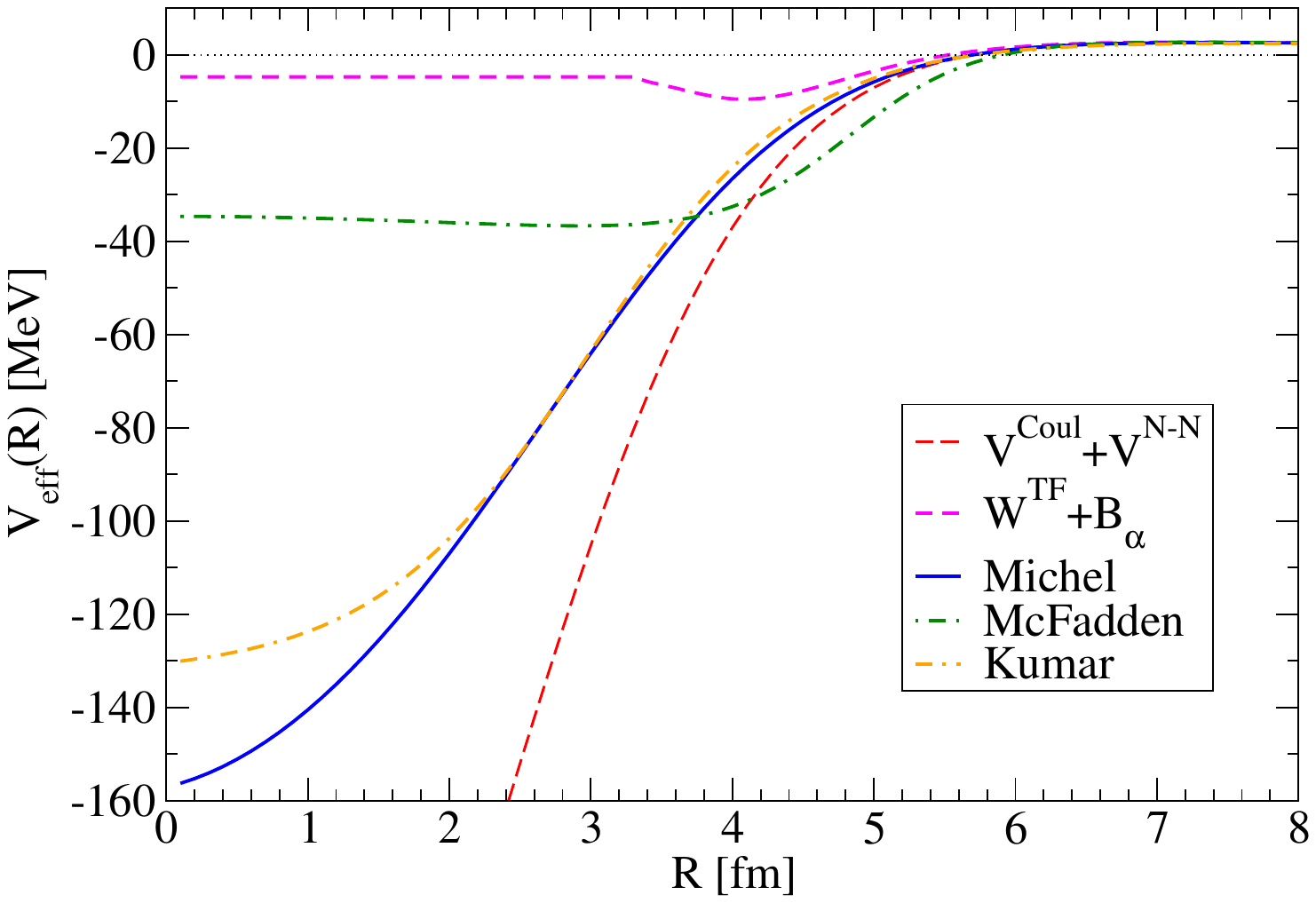}
}
  \caption{Optical model potentials from $\alpha$ - $^{16}$O scattering: The double-folding Coulomb plus nucleon-nucleon interaction and the Thomas-Fermi approach $W^{\rm TF}$ plus the medium-dependent $\alpha$-binding energy in comparison with empirical expressions by   Michel \cite{Michel}, McFadden \cite{Fadden66}, and Kumar \cite{Kumar}.}
\label{fig:W}
\end{figure}

Note that $V_{\rm eff}(R)=W( R)+B_\alpha \approx
 V^{\rm Coul}_{\alpha - {\rm O}}( R)+V^{\rm N-N}_{\alpha - {\rm O}}( R)$ 
is the mean field relative to the free $\alpha$ particle.  
Below $R = 5$ fm, Pauli blocking terms occur, see Eq. (\ref{WeffR}).
The agreement with Michel et al. \cite{Michel}  is quite good. We conclude that the choice 
(\ref{cd}) is reasonable.
The standard approaches of the optical model potentials have a diverging repulsive potential below $r_{\rm crit}$.

\section{$\alpha$-shifts}
\label{app:shifts}
{\bf This Section is not yet completed, in preparation}

In order to obtain a simple model to reproduce the essential properties of the $\alpha$-particle, we consider a microscopic model to describe correlations. With the separable interaction \cite{R1}
\begin{equation}
V(p_1,p_2;p'_1,p'_2)=- \frac{\lambda}{\Omega}e^{\frac{(p_2-p_1)^2}{4 \gamma ^2}} e^{\frac{(p'_2-p'_1)^2}{4 \gamma ^2}}\delta _{p_1+p_2,p'_1+p'_2}\delta _{\sigma \tau,\sigma' \tau'}
\end{equation}
with $\Omega$ the normalization volume, $\lambda =1449.6$ MeV fm$^3$, $\gamma = 1.152$ fm$^{-1}$, we solve the $\alpha $ cluster within a variational ansatz
\begin{equation}
\Phi_\alpha^{\rm Gauss}(p_1,p_2,p_3,p_4)=\frac{1}{\rm norm^2} e^{- (p_1^2+p_2^2+p_3^2+p_4^2) b^2/4}
\end{equation}
with the c.m. momentum $P=p_1+p_2+p_3+p_4$. The norm follows from
\begin{equation}
{\rm norm}=\sum_p e^{-b^2 p^2/2}=\int \frac{d^3p \Omega}{(2 \pi)^3} e^{-b^2 p^2/2}=\frac{\Omega}{(2 \pi b^2)^{3/2}}
\end{equation}

We calculate the kinetic energy
\begin{eqnarray}
&&T =\frac{\hbar^2}{2m}\frac{1}{\rm norm^4} \sum_{p}e^{- (p_1^2+p_2^2+p_3^2+p_4^2) b^2/2}(p_1^2+p_2^2+p_3^2+p_4^2)\nonumber \\ &&
=4 \frac{\hbar^2}{2m}\frac{1}{\rm norm}\int  \frac{d^3p \Omega}{(2 \pi)^3} e^{-b^2 p^2/2}p^2=12\frac{\hbar^2}{2mb^2}
\end{eqnarray}
From this, 1/4 is connected with the c.m. motion (introducing Jacobian coordinates, $p_1^2+p_2^2+p_3^2+p_4^2=2q_1^2+\frac{3}{2}q_2^2+\frac{4}{3}q_3^2+\frac{1}{4}P^2$). The intrinsic kinetic energy is $9\hbar^2/(2mb^2)$.

The potential energy results as
\begin{eqnarray}
\label{potalfa}
&&4^2 \frac{3}{4}\frac{1}{2}\sum_{12,1'2'} \phi(p_1)\phi(p_2)V(12,1'2')\phi(p'_1)\phi(p'_2)\nonumber \\ &&
=-6 \lambda \frac{\gamma^6 b^3}{\pi^{3/2}(\gamma^2b^2+2)^3}
\end{eqnarray}
For the total energy  the minimum -28.3087 MeV at $b = 1.93354$ occurs.
The energy pro nucleon is -7.04 MeV. The empirical rms point radius is reproduced.

In a next step, we calculate the energy per nucleon $E^{\rm free}(n_B)$ of the symmetric matter, baryon density $n_B$, in a cubic box of length $La$ with periodic boundary conditions.
The volume is $\Omega=(La)^{3}$. 
We have in the average one nucleon with given spin and isospin in the elementary box $a^3$, so that $n_B=4/a^3$.
The total number of $\alpha$-particles is $N_\alpha=L^3$, the total number of nucleons is $4 N_\alpha$.

Free nucleon states with ${\bf k} = 2\pi/(La) \{n_x,n_y,n_z\}$ are introduced, which are occupied within the Fermi cube with $k_F=\pi/a$ in all three directions $x,y,z$.
Kinetic energy is
\begin{eqnarray}
&&T_{\rm cub} =4 \times 3 \frac{\hbar^2}{2m}\int_{-\pi/a}^{\pi/a}\frac{dk_x La}{2\pi}k_x^2
\int_{-\pi/a}^{\pi/a}\frac{dk_y La}{2\pi}
\nonumber \\ &&
\times \int_{-\pi/a}^{\pi/a}\frac{dk_z La}{2\pi}=\frac{\hbar^2}{m} 2 N_\alpha \frac{\pi^2}{a^2}.
\end{eqnarray}
The KE per nucleon is $\frac{\hbar^2}{m}\frac{\pi^2}{2 \times 4^{2/3}}n_B^{2/3}$.
This value $\frac{\pi^2}{2 \times 4^{2/3}}=1.958 $ is a little bit larger than $3/10 (3 \pi^2/2)^{2/3}=1.808$ for the Fermi sphere instead of the Fermi cube.

The potential energy is
\begin{eqnarray}
&&V_{\rm cub}=4\times 3/2 \sum_{12,1'2'}V(12,1'2')=-6 \frac{\lambda}{\Omega}
\int_{-\pi/a}^{\pi/a}\frac{dk_1^x La}{2\pi}
\nonumber \\ &&
\dots \int_{-\pi/a}^{\pi/a}\frac{dk_2^z La}{2\pi} e^{-(k_2^x-k_1^x)^2/2 \gamma^2}\dots e^{-(k_2^z-k_1^z)^2/2 \gamma^2}.
\end{eqnarray}
With
\begin{eqnarray}
&&I=\int_{-\pi/a}^{\pi/a}dk_1^x \int_{-\pi/a}^{\pi/a}dk_2^x e^{-(k_2^x-k_1^x)^2/2 \gamma^2}
\nonumber \\ &&
=2 \gamma\left[\left(e^{-2 \pi^2/a^2 \gamma^2}-1\right) \gamma+\frac{1}{a}2^{1/2} \pi^{3/2} {\rm erf}(2^{1/2}\pi/a \gamma)\right]
\nonumber \\ &&
{}
\end{eqnarray}
so that
\begin{equation}
V_{\rm cub}=-6 \frac{\lambda \Omega}{(2 \pi)^6} I^3,\qquad a=(4/n_B)^{1/3}.
\end{equation}
and the potential energy per nucleon $-6 \frac{\lambda}{(2 \pi)^6 n_B} I^3$. The energy per nucleon comes out as
\begin{equation}
\label{SEsep}
E_{\rm cub}(n_B)=\frac{\hbar^2}{m} \frac{\pi^2}{2\times 4^{2/3}} n_B^{2/3}-6 \frac{\lambda}{(2 \pi)^6 n_B} I^3.
\end{equation}

We compare this result with standard expressions.
The chemical potential contains the kinetic energy {degeneracy 4, Fermi wave number $k_F=(3 \pi^2 n/2)^{1/3}$)
\begin{equation}
E_{\rm kin,Fermi}(n)=\frac{\hbar^2}{2 m}k_F^2=\frac{\hbar^2}{2 m}\left(\frac{3 \pi^2}{2}\right)^{2/3} n^{2/3}
\end{equation}
so that $\mu(n)=E_{\rm kin,Fermi}(n)+\Delta E^{\rm SE}(n)$.
The self-energy shift of the single-nucleon states can be estimated by the Skyrme model,  
\begin{equation}
\Delta E^{\rm SE}(n)=- \frac{3}{4}1057.3 n+\frac{3}{16} 14463.5 n^2
\end{equation}
The energy per nucleon follows as 
\begin{eqnarray}
&&E/N=\frac{1}{n} \int_0^n \mu(n') dn'  \\ &&=\frac{3 \hbar^2}{10 m}\left(\frac{3 \pi^2}{2}\right)^{2/3} n^{2/3} -\frac{3}{8}1057.3 n+\frac{1}{16} 14463.5 n^2.\nonumber
\end{eqnarray}
A better parametrization is given by the RMF approach, the DD2 version gives
\begin{equation}
\mu(n)=((m c^2-s(n))^2+(\hbar c k_F)^2)^{1/2}-m c^2+v(n)
\end{equation}
(for a parametrization of $s$ and $v$ see \cite{R}.)
The minimum occurs at $E( 0.148327 {\rm fm}^{-3})=-16.2784$ MeV.
In the subsaturated range of density, the three approaches are in reasonable agreement, see Fig. \ref{fig:SE}.

\begin{figure}
\resizebox{0.5\textwidth}{!}{%
  \includegraphics{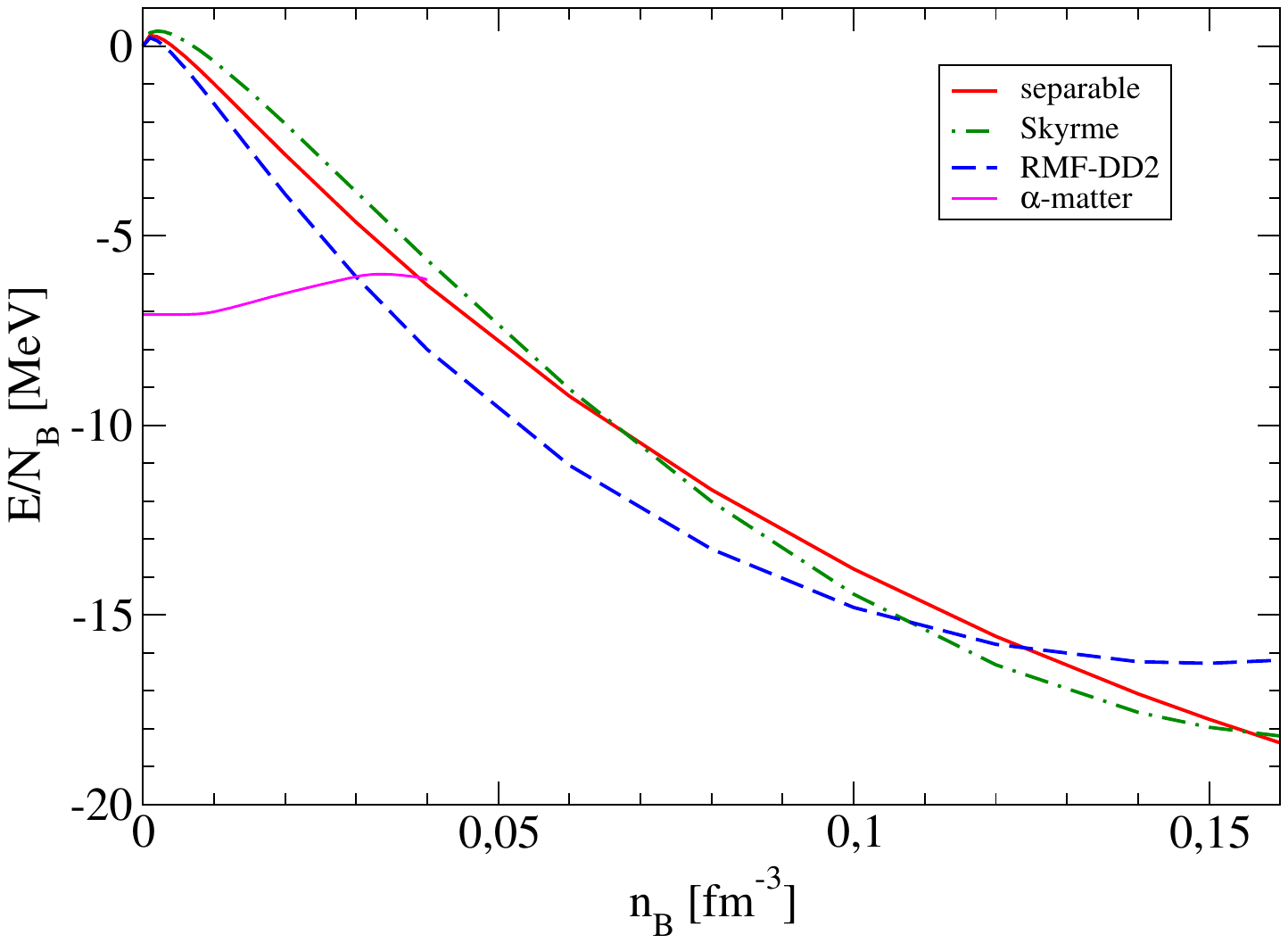}
}
  \caption{Energy per nucleon as function of the nucleon density $n_B$. Results for the separable potential (\ref{SEsep}) are compared with RMF-DD2 and Skyrme calculations. The energy  per nucleon for $\alpha$-matter (magenta full line) takes at zero density the value -28.3/4 MeV. }
\label{fig:SE}
\end{figure}

Between both limits, the free $\alpha$-cluster gas at low densities and the free-nucleon quasiparticle gas at high densities, we consider a Bloch ansatz \cite{Tak04}.
\begin{equation}
\phi_k(p)=\frac{1}{N_k^{1/2}N_\alpha^{1/2}}(2 \pi b^2)^{1/4}\sum_{m=-N_\alpha/2}^{N_\alpha/2}e^{imka+impa-b^2p^2/4}
\end{equation}

The kinetic energy follows as (4 for spin/isospin, 3 for the components $x,y,z$)
\begin{eqnarray}
&&T_{\rm Bloch}=4\times 3 \frac{\hbar^2}{2m} \sum_k \nonumber \\ && \times \frac{\sum_{m_1m_2} \sum_p e^{i(m_1-m_2)(k^x+p)a-b^2p^2/2} p^2} { \sum_{m_1m_2} \sum_{p'} e^{i(m_1-m_2)(k+{p'})a-b^2{p'}^2/2}}.
\end{eqnarray}
After performing the $p,p'$ integrals we have
\begin{eqnarray}
&&T_{\rm Bloch}/N_B= 3 \frac{\hbar^2}{2mb^2}\int_{-\pi/a}^{\pi/a} \frac{dk^x a}{2 \pi}\nonumber \\&& \times \frac{ \sum_{m=-L/2}^{L/2} e^{imk^xa-a^2 m^2/(2 b^2)}\left(1-\frac{m^2 a^2}{b^2}\right)}{ \sum_{m'=-L/2}^{L/2} e^{im'ka-a^2 m'^2/(2 b^2)}}.
\end{eqnarray}
This expression contains the c.m. kinetic energy.  
%
%
%
To separate the c.m. energy, it was proposes in \cite{Tak04} to consider the ratio $(x=a/b)$
\begin{equation}
\label{kinnorm}
N_c(x)= 1-\frac{1}{4}\left[1-\frac{\sum_{m=-L/2}^{L/2} e^{-x^2 m^2/2}(m^2 x^2)}
{ \sum_{m'=-L/2}^{L/2} e^{-x^2 m'^2/2}}\right]
\end{equation}
as a factor for the kinetic energy to exclude the c.m. kinetic energy in the low-density region.
%
The evaluation of the potential energy is somewhat lengthy so that  we give only the final result
\begin{eqnarray}
&&V_{\rm Bloch}=   -\frac{3}{2 \pi^{3/2}}\frac{\lambda b^3}{(b^2+2/\gamma^2)^3} \left(\int_{-\pi/a}^{\pi/a}\frac{dk_1 a}{2 \pi} \int_{-\pi/a}^{\pi/a}\frac{dk_2 a}{2 \pi} \right.\nonumber \\ &&\left.
\times \frac{\Sigma_1^e[\Sigma_2^e\Sigma_3^e+\Sigma_2^o\Sigma_3^o]+\Sigma_1^o[\Sigma_2^e\Sigma_3^o+\Sigma_2^o\Sigma_3^e]}{\Sigma_4(k_1) \Sigma_4(k_2)}\right)^3
\end{eqnarray}
where
\begin{eqnarray}
 &&\Sigma_1^e=\sum_m e^{i 2m(k_1+k_2)a/2-m^2x^2}   \nonumber \\
 &&\Sigma_1^o=\sum_m e^{i (2m+1)(k_1+k_2)a/2-(2m+1)^2x^2/4} \nonumber \\
 &&\Sigma_2^e=\sum_m e^{i 2m(k_2-k_1)a/2-2 m^2a^2/(b^2+2/\gamma^2)}   \nonumber \\
&&\Sigma_2^o=\sum_m e^{i (2m+1)(k_2-k_1)a/2-(2 m+1)^2a^2/(2b^2+4/\gamma^2)}   \nonumber \\
&&\Sigma_3^e=\sum_m e^{i 2m(k_1-k_2)a/2-2 m^2a^2/(b^2+2/\gamma^2)}   \nonumber \\
&&\Sigma_3^o=\sum_m e^{i (2m+1)(k_1-k_2)a/2-(2 m+1)^2a^2/(2b^2+4/\gamma^2)}   \nonumber \\
&&\Sigma_4(k)=\sum_m e^{i mka-m^2a^2/(2 b^2)}   \nonumber \\
\end{eqnarray}

Within our approach one has to search for the minimum of the energy as function of the width parameter $b$, see Fig. \ref{fig:bsep}. 
At low densities, the minimum occurs at $b=1.934$ fm. This value is slightly increasing with increasing density. 
At the nucleon density $n_B=0.0387$ fm$^{-3}$ it jumps to the free nucleon value, see Fig. \ref{fig:bsep}. 
This is a sharp transition. 
It is not clear whether this phase transition is due to the approximations such as the separation of the c.m. kinetic energy or the Gauss ansatz for the wave function, or is a real sharp quantum phase transition to a correlated state. To understand the dissolution of the bound state in the case of a sharp quantum phase transition, we can use the equilibrium condition of equal chemical potential $\mu$ in both phases. With the energy per nucleon $e(n_B)=E/N_B$ we have
\begin{equation}
    \mu(n_B)=e(n_B)+n_B \frac{\partial e(n_B)}{\partial n_B}.
\end{equation}
The disappearance of the $\alpha$-matter phase occurs if the chemical potential coincides with the free momentum quasiparticle phase.
\\

\begin{figure}
\resizebox{0.5\textwidth}{!}{%
  \includegraphics{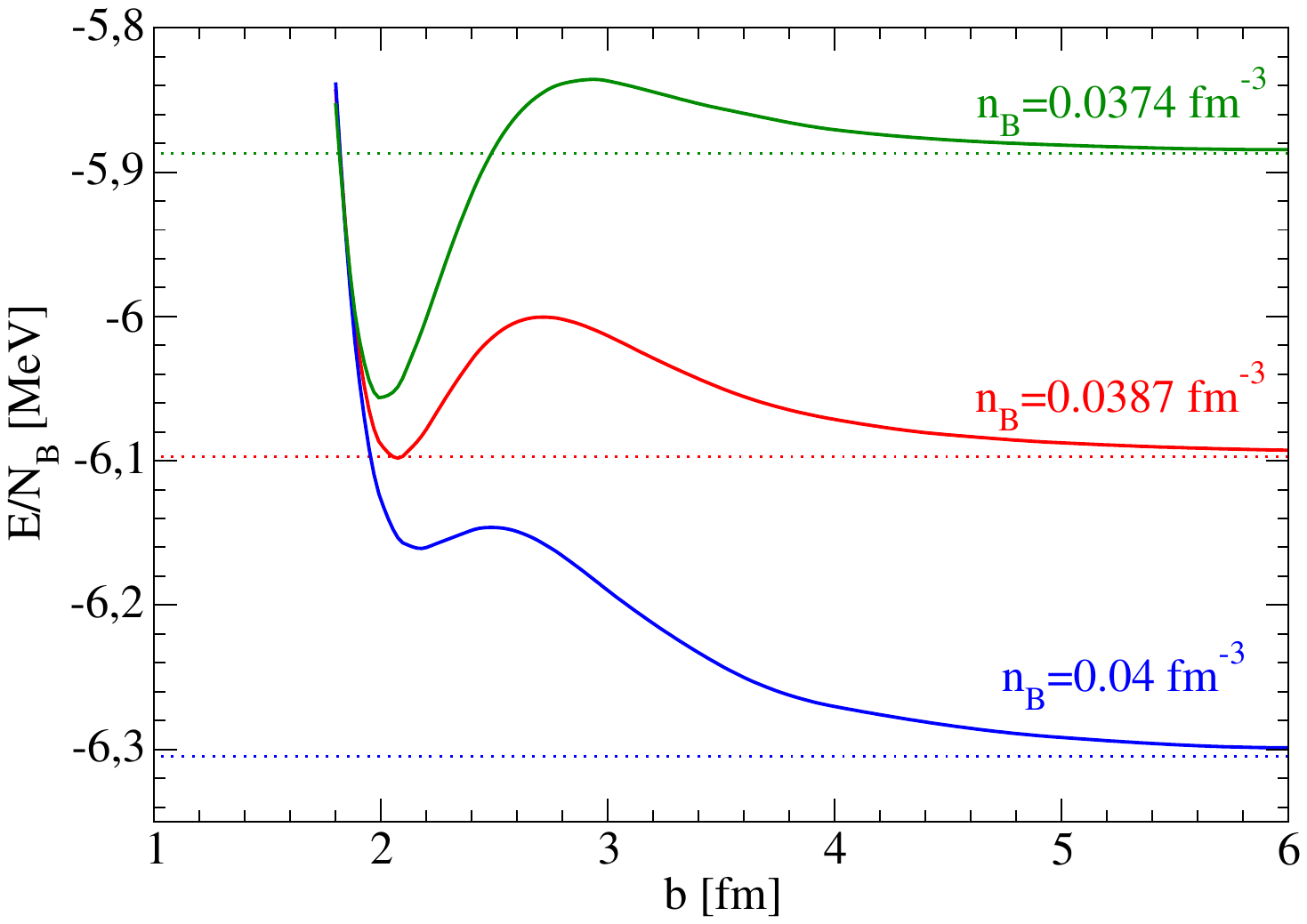}
}
  \caption{Energy per nucleon as function of the width parameter $b$ at different nucleon density $n_B$. 
  The result (\ref{SEsep}) for the uncorrelated nuclear matter is shown as dotted line.}
\label{fig:bsep}
\end{figure}

 }

%


\begin{thebibliography}{99}


\bibitem{RingSchuck}
P. Ring and P. Schuck,
{\it The Nuclear Many-Body Problem} (Springer, Berlin 1980).

\bibitem{Po14}
G. R\"opke, P. Schuck, Y. Funaki, H. Horiuchi, Zhongzhou Ren, A. Tohsaki, Chang Xu, T. Yamada, and Bo Zhou,
 Phys. Rev. C {\bf 90}, 034304 (2014).

\bibitem{Xu16}
Chang Xu, Zhongzhou Ren, G. R\"opke, P. Schuck, Y. Funaki, H. Horiuchi, A. Tohsaki, T. Yamada, and Bo Zhou,
Phys. Rev. C {\bf 93}, 011306({}R) (2016).

\bibitem{Xu17}
Chang Xu, G. R\"opke, P. Schuck, Zhongzhou Ren, Y. Funaki, H. Horiuchi, A. Tohsaki, T. Yamada, and Bo Zhou, Phys. Rev. C {\bf 95}, 061306({}R) (2017).

 \bibitem{Yang20}
Shuo Yang, Chang Xu, Gerd R\"opke, Peter Schuck, Zhongzhou Ren, Yasuro Funaki, Hisashi Horiuchi, Akihiro Tohsaki, Taiichi Yamada, and Bo Zhou, Phys. Rev. C {\bf 101}, 024316 (2020).

 \bibitem{Yang21}
Shuo Yang, Chang Xu, and Gerd R\"opke, Phys. Rev. C {\bf 104}, 034302 (2021).

 \bibitem{Bai19}
Dong Bai, Zhongzhou Ren, and Gerd R\"opke, Phys. Rev. C {\bf 99}, 034305 (2019).

 \bibitem{Wang22}
Zhen Wang, Dong Bai, and Zhongzhou Ren, Phys. Rev. C {\bf 105}, 024327 (2022).

\bibitem{Jin23}
Zisheng Jin, Mingshuai Yan, Hao Zhou, An Cheng, Zhongzhou Ren, and Jian Liu, Phys. Rev. C {\bf 108}, 014326 (2023).

\bibitem{Li23}
Ruijia Li and Chang Xu, Phys. Rev. C {\bf 107}, 064301 (2023).

 \bibitem{Yang23}
Shuo Yang, Ruijia Li, and Chang Xu, Phys. Rev. C {\bf 108}, L021303 (2023).

\bibitem{THSR} 
A. Tohsaki, H. Horiuchi, P. Schuck, and G. R\"opke, 
Phys. Rev. Lett. {\bf 87}, 192501 (2001).

\bibitem{Bo12}
Bo Zhou, Zhongzhou Ren, Chang Xu, Y. Funaki, T. Yamada, A.
Tohsaki, H. Horiuchi, P. Schuck, and G. R\"opke, Phys. Rev. C {\bf
86}, 014301 (2012).

\bibitem{Bo13}
Bo Zhou, Y. Funaki, H. Horiuchi, Zhongzhou Ren, G. R\"opke, P. Schuck, A. Tohsaki,
Chang Xu, and T. Yamada,
Phys. Rev. Lett. {\bf 110}, 262501 (2013).

\bibitem{Bo14}
Bo Zhou, Y. Funaki, H. Horiuchi, Zhongzhou Ren, G. R\"opke, P.
Schuck, A. Tohsaki, Chang Xu, and T. Yamada, Phys. Rev. C {\bf
89}, 034319 (2014).

\bibitem{Lyu}
Mengjiao Lyu, Zhongzhou Ren, Bo Zhou, Yasuro Funaki, Hisashi Horiuchi, Gerd R\"opke, Peter Schuck, Akihiro Tohsaki, Chang Xu, and Taiichi Yamada, Phys. Rev. C {\bf 91}, 014313 (2015).

\bibitem{Lyu1}
Mengjiao Lyu, Zhongzhou Ren, Bo Zhou, Yasuro Funaki, Hisashi Horiuchi, Gerd R\"opke, Peter Schuck, Akihiro Tohsaki, Chang Xu, and Taiichi Yamada,
Phys. Rev. C {\bf 93}, 054308 (2016).

\bibitem{Bo23}
Bo Zhou, Yasuro Funaki, Hisashi Horiuchi, Yu-Gang Ma, Gerd R\"opke, Peter Schuck, Akihiro Tohsaki, and Taiichi Yamada, Nature Comm. {\bf 14} (2023) 8206.

\bibitem{Freer18}
M. Freer, H. Horiuchi, Y. Kanada-En’yo, D. Lee, and U.-G. Meißner
Rev. Mod. Phys. {\bf 90}, 035004 (2018). 

 \bibitem{wir}
G. R\"opke, P. Schuck, Chang Xu, Zhongzhou Ren, M. Lyu,  Bo Zhou, Y. Funaki, H. Horiuchi, A. Tohsaki,  and T. Yamada, 
J. Low. Temp. Phys. {\bf 189}, 383 (2017) [arXiv:1707.04517 [nucl-th]].

\bibitem{R17}
G. R\"opke, P. Schuck, Y. Funaki, H. Horiuchi, Zhongzhou Ren, A. Tohsaki, Chang Xu, T. Yamada, and Bo Zhou,
 J. Phys. Conf. Ser. {\bf 863}, 012006 (2017).
 
 \bibitem{Ro18}
G. R\"opke,, in {\it  Proceedings of the 4th International Workshop on "State of the Art in Nuclear Cluster  Physics" (SOTANCP4)},
AIP Conf. Proc. 2038, 020008-1-020008-10 (AIP, New York, 2018).
arXiv:1810.01274 [nucl-th].


\bibitem{Qu2011}
W.W. Qu, G.L. Zhang, X.Y. Le, Nucl. Phys. A {\bf 868} 1 (2011).

\bibitem{M3YReview}
G.R. Satchler and  W.G. Love, Phys. Rep. {\bf 55}, 183 (1979).

\bibitem{Mirea}
M. Mirea, 
Phys. Rev. C {\bf 96}, 064607 (2017).

\bibitem{WS}
N. Schwierz, I. Wiedenhover, and A. Volya, {\it Parameterization of the Woods-Saxon Potential for Shell-Model Calculations}. arXiv:0709.3525 (2007).

\bibitem{Tak04}
Hiroki Takemoto, Masahiro Fukushima, Satoshi Chiba, Hisashi Horiuchi, Yoshinori Akaishi, and Akihiro Tohsaki, Phys. Rev. C {\bf 69}, 035802 (2004).

\bibitem{Mat75}
T. Matsuse, M. Kamimura, and Y. Fukushima, Prog. Theor. Phys. {\bf 53}, 706 (1975).

\bibitem{Fun08}
Y. Funaki, H. Horiuchi, G. R\"opke, P. Schuck, A. Tohsaki, and T. Yamada, Phys. Rev. C {\bf 77}, 064312 (2008).

\bibitem{Ebran20}
J.-P. Ebran, M. Girod, E. Khan, R. D. Lasseri, and P. Schuck,
Phys. Rev. C {\bf 102}, 014305 (2020).

\bibitem{R1}
G. R\"opke, Phys. Rev. C {\bf 79}, 014002 (2009).

\bibitem{R11}
G. R\"opke, 
Nucl. Phys. A {\bf 867}, 66 (2011).

\bibitem{T14}
S. Typel,
Phys. Rev. C {\bf 89}, 064321 (2014).

\bibitem{T20}
Z. Yang, J. Tanaka, S. Typel, T. Aumann, J.
Zenihiro, S. Adachi, S. Bai, P. van Beek, D. Beaume,
Y. Fujikawa, J. Han, S. Heil, S. Huang, A. Inoue,
Y. Jiang, M. Kn\"osel, N. Kobayashi, Y. Kubota, W. Liu,
J. Lou, Y. Maeda, Y. Matsuda, K. Miki, S.
Nakamura, K. Ogata, V. Panin, H. Scheit, F.
Schindler, P. Schrock, D. Symochko, A. Tamii, T.
Uesaka, V. Wagner, K. Yoshida,
JPS Conf. Proc. {\bf 31}, 011019 (2020).

\bibitem{T21}
J. Tanaka,
Z.H. Yang,
S. Typel,
S. Adachi,
S. Bai,
P. Van Beek,
D. Beaumel,
Y. Fujikawa,
J. Han,
S. Heil,
S. Huang,
A. Inoue,
Y. Jiang,
M. Kn\"osel,
N. Kobayashi,
Y. Kubota,
W. Liu,
J. Lou,
Y. Maeda,
Y. Matsuda,
K. Miki,
S. Nakamura,
K. Ogata,
V. Panin,
H. Scheit,
F. Schindler,
P. Schrock,
D. Symochko,
A. Tamii,
T. Uesaka,
V. Wagner,
K. Yoshida,
J. Zenihiro,
T. Aumann,
Science  {\bf 371}, 200 (2021).

\bibitem{AMD} 
Y. Kanada-En'yo, Prog.. Theor. Phys. {\bf 117}, 655 (2007).

\bibitem{AMD04}
A. Ono, H. Horiuchi, Prog. Part. Nucl. Phys. {\bf 53} (2004) 501.

\bibitem{Yos19}
K. Yoshida, Y. Chiba, M. Kimura, Y. Taniguchi, Y. Kanada-En'yo, and K. Ogata, Phys. Rev. C {\bf 100}, 044601 (2019).

\bibitem{Chi17}
Y. Chiba and M. Kimura, Prog. Theor. Exp. Phys. {\bf 2017}, 053D01 (2017). 

\bibitem{Tan21}
Y. Taniguchi, K. Yoshida, Y. Chiba, Y. Kanada-En'yo, M. Kimura, and K. Ogata, PRC {\bf 103}, L031305 (2021).

\bibitem{Yos22}
K. Yoshida and J. Tanaka, Phys. Rev. C {\bf 106}, 014621 (2022).

\bibitem{Qi10}
C. Qi, A.N. Andreyev, M. Huyse, R.J. Liotta, P. Van Duppen, 
and R.A. Wyss, Phys. Rev. C {\bf 81}, 064319 (2010).

\bibitem{Nak23}
T. Nakatsukasa, N. Hinohara, Phys. Rev. C {\bf 108}, 014318 (2023).

\bibitem{Kimura22}
Q. Zhao, M. Kimura, B. Zhou, and S. Shin,
Phys. Rev. C {\bf 106}, 054313 (2022).

\bibitem{PG11}
P.-G. Reinhard, J. A. Maruhn, A. S. Umar, and V. E. Oberacker, Phys. Rev. C {\bf 83}, 034312 (2011).

\bibitem{Toh17}
A. Tohsaki, H. Horiuchi, P. Schuck, and G. R\"opke, 
Rev. Mod. Phys. {\bf 89}, 011002 (2017).

\bibitem{Bo}
Bo Zhou, Zhongzhou Ren, Chang Xu, Y. Funaki, T. Yamada, A.
Tohsaki, H. Horiuchi, P. Schuck, and G. R\"opke, Phys. Rev. C {\bf
86}, 014301 (2012).

\bibitem{Bo2}
Bo Zhou, Y. Funaki, H. Horiuchi, Zhongzhou Ren, G. R\"opke, P.
Schuck, A. Tohsaki, Chang Xu, and T. Yamada, Phys. Rev. C {\bf
89}, 034319 (2014).

\bibitem{Bo3}
Bo Zhou, Y. Funaki, H. Horiuchi, Zhongzhou Ren, G. R\"opke, P. Schuck, A. Tohsaki,
Chang Xu, and T. Yamada,
Phys. Rev. Lett. {\bf 110}, 262501 (2013).

\bibitem{Fadden66}
L. McFadden and G.R. Satchler,
Nucl. Phys. {\bf 84},  177 (1966).

\bibitem{Michel}
F. Michel, J. Albinski, P. Belery, T. Delbar, G. Gr\'egoire, B. Tasiaux, and G. Reidemeister, 
Phys. Rev. C {\bf 28}, 1904 (1983).

\bibitem{Oertzen}
W. von Oertzen, M. Freer, and Y. Kanada-Enyo, Phys. Reports {\bf 432}, 43(2006);

\bibitem{Oertzen1}
W. von Oertzen, in: C. Beck (Ed.), {\it Clusters in Nuclei}, vol. 1, p.109 (Springer, Berlin 2010).

\bibitem{Ohkubo}
Y. Hirabayashi and S. Ohkubo,
Phys. Rev. C {\bf 88}, 014314 (2013).

\bibitem{Fukui16}
T. Fukui, Y. Taniguchi, T. Suhara, Y. Kanada-En'yo, and K. Ogata, 
 Phys. Rev. C {\bf 93}, 034606 (2016).

\bibitem{Kumar}
Ashok Kumar and S. Kailas,
Nucl. Phys. A {\bf 776},  105 (2006).

\bibitem{R}
G. R\"opke, 
 Phys. Rev. C {\bf 92}, 054001 (2015).

\end{thebibliography}
\end{document}